\documentclass[aps,apl,reprint,superscriptaddress,longbibliography]{revtex4-2}

\usepackage{graphicx}
\usepackage{amsmath,amsfonts,amssymb}	
\usepackage{mathrsfs}
\usepackage[normalem]{ulem}
\usepackage[dvipsnames]{xcolor}
\usepackage{epstopdf}
\usepackage[retain-explicit-plus]{siunitx}
\usepackage[version=4]{mhchem}
\usepackage{graphicx}
\usepackage{array}
\usepackage{tabularx}
\newcolumntype{Y}{>{\centering\arraybackslash}X}
\usepackage[section]{placeins}

\usepackage{longtable}
\usepackage{braket}
\usepackage{booktabs}
\usepackage{hyperref}
\usepackage{cleveref}
\usepackage{tabularx}
\usepackage{makecell}

\hypersetup{
    colorlinks=true,
    linkcolor=blue,
    filecolor=magenta,      
    urlcolor=blue,
    citecolor=blue
    }
\definecolor{alizarin}{rgb}{0.82, 0.1, 0.26} 
\definecolor{violet}{rgb}{0.5, 0.1, 1} 
\definecolor{darkgreen}{rgb}{0, 0.8, 0} 

\usepackage{bibunits}
\defaultbibliographystyle{apsrev4-2}
\defaultbibliography{bib}

\draft 

\begin{document}

\title{Rapid estimation of synthesizability windows of inorganic materials from first principles}


\author{Finja Tadge}
\email{finta@dtu.dk}
\affiliation{National Centre for Nano Fabrication and Characterization (DTU Nanolab), Technical University of Denmark, 2800 Kongens Lyngby, Denmark}

\author{Javier Sanz Rodrigo}
\affiliation{National Centre for Nano Fabrication and Characterization (DTU Nanolab), Technical University of Denmark, 2800 Kongens Lyngby, Denmark}

\author{Andrea Crovetto}%
\email{ancro@dtu.dk}
\affiliation{National Centre for Nano Fabrication and Characterization (DTU Nanolab), Technical University of Denmark, 2800 Kongens Lyngby, Denmark}


\begin{abstract}
Fast prediction of the synthesizability conditions of materials remains challenging, even assuming synthesis under thermodynamic equilibrium. We combine density functional theory (DFT) with machine-learned interatomic potentials to enable high-throughput generation of phase predominance diagrams as a function of temperature and partial pressures of the gaseous reactants. These diagrams can immediately be used by experimentalists to translate computational predictions into real synthesis parameters in the lab. Predominance diagrams are generated for a diverse set of binary compounds and for 48 more complex ternary metal phosphosulfide systems, but the method is in principle scalable to any inorganic material class. The calculated predominance diagrams generally show good agreement with the experimental synthesis literature, with a drastic reduction in computational cost compared to a full DFT approach. We find several examples of compounds that appear as metastable in a zero-temperature stability hull picture, but that become thermodynamically stable under well-defined synthesis windows. 

\end{abstract}

\maketitle

\section{Introduction}

The energy above the zero-temperature convex stability hull $E_h$ is often employed as a metric for assessing the synthesizability of compounds~\cite{MP, kirklinOpenQuantumMaterials2015}. While $E_h$ has traditionally been calculated with density functional theory (DFT), neural network models have now extended the assessment of thermodynamic stability to the range of $10^9$ materials.~\cite{schmidtMachineLearningAssistedDeterminationGlobal2023}. However, such zero-temperature approaches do not necessarily reflect realistic temperature- and pressure-dependent synthesis conditions.

Accurate computational phase diagrams at finite temperatures and pressures require explicit modeling of the temperature-dependent heat capacity of materials and of the entropic contributions to their Gibbs free energy. Phonon calculations are needed to determine heat capacities and vibrational entropy~\cite{Ba_Zr_S_phase_diagrams}. Total energy calculations on a large number of structures generated by quasi-random~\cite{zungerSpecialQuasirandomStructures1990} or Monte Carlo sampling~\cite{configurational_entropy_2} are commonly employed to estimate configurational entropy. DFT-based versions of these methods are still too computationally expensive for large-scale, high-throughput phase diagram prediction.

Less costly alternatives, including machine learning–based descriptor approaches for Gibbs free energy prediction, have been proposed~\cite{gibbs_descriptor}. While useful, descriptor-based methods are intrinsically heuristic and may not generalize well to material systems beyond the training set. Similarly, machine learning models trained on text-mined synthesis recipes have so far shown limited generalizability for predicting experimental synthesis conditions~\cite{sunCriticalReflectionAttempts2025} and a tendency to overestimate the likelihood of synthesis~\cite{schlesingerThermodynamicAssessmentMachine2026}.
Consequently, large-scale, high-throughput synthesizability prediction across complex phase spaces under realistic experimental conditions remains a major challenge.

With the recent emergence of machine-learned interatomic potentials (MLIPs), the first low-cost workflows to integrate physics-based entropy contributions to the Gibbs free energy of materials have been proposed~\cite{tolborgLowCostVibrationalFree2023,zhuMachineLearningPotentials2025,unglertActiveLearningPotentials2026}.
These initial reports have focused on systems where configurational entropy is dominant, such as solid solutions of metals and of gapped solids~\cite{tolborgLowCostVibrationalFree2023} and high-entropy alloys~\cite{zhuMachineLearningPotentials2025}. There are still at least three open questions. 

First, can MLIP-based methods produce high-throughput phase diagrams of nominally ordered inorganic materials with sufficient quality to inform experimental synthesis? For these materials it is important to include vibrational energy and heat capacity, but it may be acceptable to neglect configurational entropy, which is more costly to calculate even with MLIPs~\cite{zhuMachineLearningPotentials2025}.

Second, can the latest generation of universal MLIP foundation models significantly improve predictive power in materials thermochemistry over their predecessors? In particular, the MatterSim MLIP foundation model~\cite{MatterSim} has very recently been shown to perform particularly well at phonon calculations in a benchmark study~\cite{loewUniversalMachineLearning2025}. Due to its accuracy at predicting phonon frequencies and the phonon density of states (DOS), MatterSim gives mean absolute errors (per unit cell) of only 15~J/(mol$\cdot$K) in vibrational entropy and 3~J/(mol$\cdot$K) in heat capacity at constant volume with respect to explicit DFT calculations at the PBE level.

\begin{figure}[]
    \centering
    \includegraphics[width=0.8\linewidth]{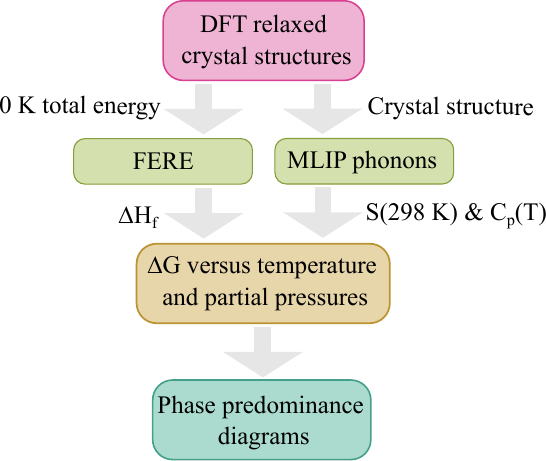}
    \caption{Workflow for high-throughput calculation of thermodynamic stability windows of materials versus temperature and partial pressures. Phase predominance diagrams are constructed based on the formation enthalpy $\Delta H_f$, vibrational entropy $S$(298~K) and heat capacity $C_p(T)$ calculated by combining DFT with the FERE approach and MLIP phonon calculations.}
    \label{fig:workflow}
\end{figure}

Third, can high-throughput calculated phase diagrams be presented in a way that is intuitively understood by experimental synthesis scientists? Although MLIP-generated phase diagrams for metallic solid solutions have been plotted as a function of temperature and metal composition~\cite{tolborgLowCostVibrationalFree2023,zhuMachineLearningPotentials2025}, calculated phase diagrams of compounds have traditionally been plotted as a function of temperature and atomic chemical potentials~\cite{gibbs_descriptor}. From the perspective of an experimentalist, synthesizability windows of materials are easiest to translate into real synthesis processes when visualized as a function of reactive gas partial pressures rather than chemical potentials~\cite{BaZrS_phase_diagrams}.

In this work we address the questions above.
In a first step (Fig.~\ref{fig:workflow}), DFT calculations are used to compute enthalpies at zero temperature, with the accuracy of formation energies improved using the fitted elemental reference energies (FERE) approach~\cite{FERE_paper}. Subsequently, the MatterSim MLIP foundation model~\cite{MatterSim} is employed for rapid phonon band structure calculations to determine vibrational entropies and heat capacities. This calculation step takes just a few minutes on a standard compute node, thus circumventing the bottleneck of DFT phonon calculations in computational thermochemistry. From this information, the Gibbs free energies of all compounds, are computed as a function of temperature and gas-phase chemical potentials using standard thermochemical relationships. The resulting phase predominance diagrams can be conveniently represented in two dimensions of choice for intuitive understanding by experimentalists looking to synthesize the materials. Combining ab-initio DFT calculations, MLIPs and standard thermochemistry provides a physics-based (rather than heuristic) workflow extendable to different material classes across the periodic table.

\begin{figure}[]
    \centering
    \includegraphics[width=\columnwidth]{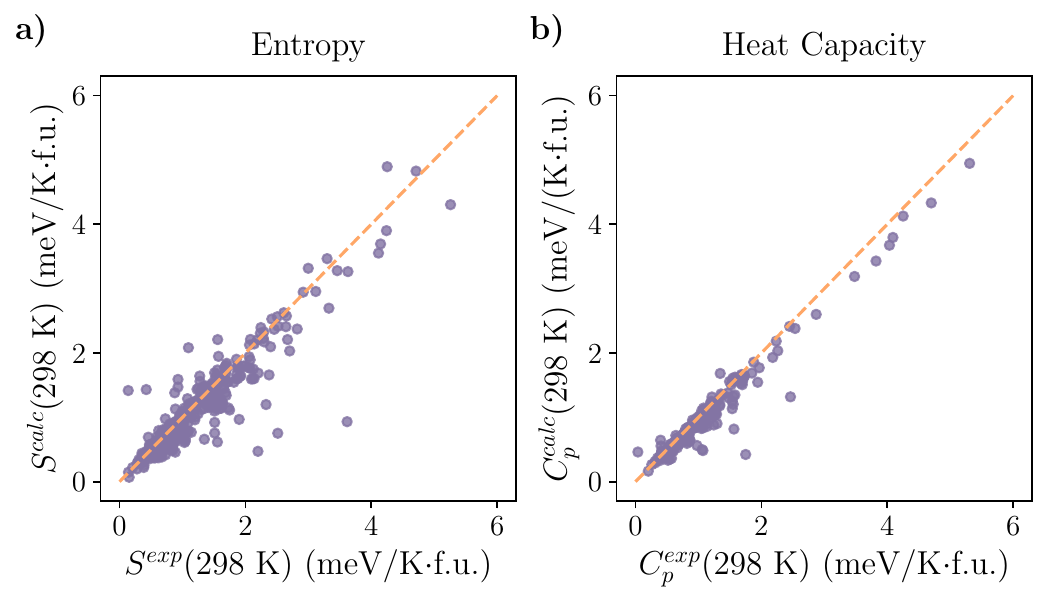}
    \caption{MLIP-calculated values of $S$(298~K) and $C_{p}$(298~K) in meV/(K$\cdot$formula unit) compared to experimental values for all binaries considered in the different datasets.}
    \label{fig:Comparison_S_Cp}
\end{figure}

We evaluate the chemical versatility of this workflow by generating predominance diagrams in four chemically diverse binary systems (oxides, nitrides, sulfides, phosphides) and comparing them to experiment. As a high-throughput application of this method, we generate phase diagrams for 48 distinct metal phosphosulfide ternary systems, comprising over 1000 ternary compounds and about 450 binaries, including 19 new materials that we recently found to lie on the 0~K stability hull~\cite{Javier_goat}. The predicted windows of thermodynamic stability show promising agreement with available experimental synthesis conditions for bulk materials, highlighting the potential of this approach for high-throughput prediction of synthesis conditions of computationally discovered materials.

\section{Results} \label{sec:results}

\subsection{Binary systems: Oxides, nitrides, sulfides, phosphides} 
\label{sec:predominance_diagrams}

\begin{table*}[]
\centering
\begin{tabular}{c | c c c c c c }
    Dataset \hspace*{3mm} & \hspace*{3mm} N \hspace*{3mm} & \hspace*{3mm} n$_{el}$ \hspace*{3mm} & \hspace*{3mm} \makecell{MAE $\Delta H_f$ \\ (eV/atom)} \hspace*{3mm} & \hspace*{3mm} \makecell{MAE $\Delta H_f^{FERE}$ \\ (eV/atom)} \hspace*{3mm} \\ 
    \hline \hline
    Binary oxides & 150 & 74 & 0.236 & 0.069 \\ \hline 
    Binary sulfides & 118 & 68 & 0.204 & 0.058 \\ \hline
    Binary nitrides and phosphides  & 98 & 53 & 0.160 & 0.055\\ \hline
    \makecell{Binary and ternary \\ phosphides and sulfides} \hspace*{3mm} & 144 & 50 & 0.152 & 0.073 \\
    \hline \hline
\end{tabular}
\caption{Statistics of elemental reference-phase energy corrections for four different datasets. Reference energies for n$_{el}$ elements are simultaneously fitted against the formation enthalpy of N distinct compounds containing those elements. MAEs for the formation enthalpy before ($\Delta H_f$) and after ($\Delta H_f^{FERE}$) fitting are reported. Corrections from the first three rows are used to plot the binary-system predominance diagrams presented in this work. Corrections from the last row are used to plot predominance diagrams for ternary phosphosulfide systems.}
\label{tab:FERE_MAE_different_datasets}
\end{table*}

\begin{figure*}[]
    \centering
    \includegraphics[width=\linewidth]{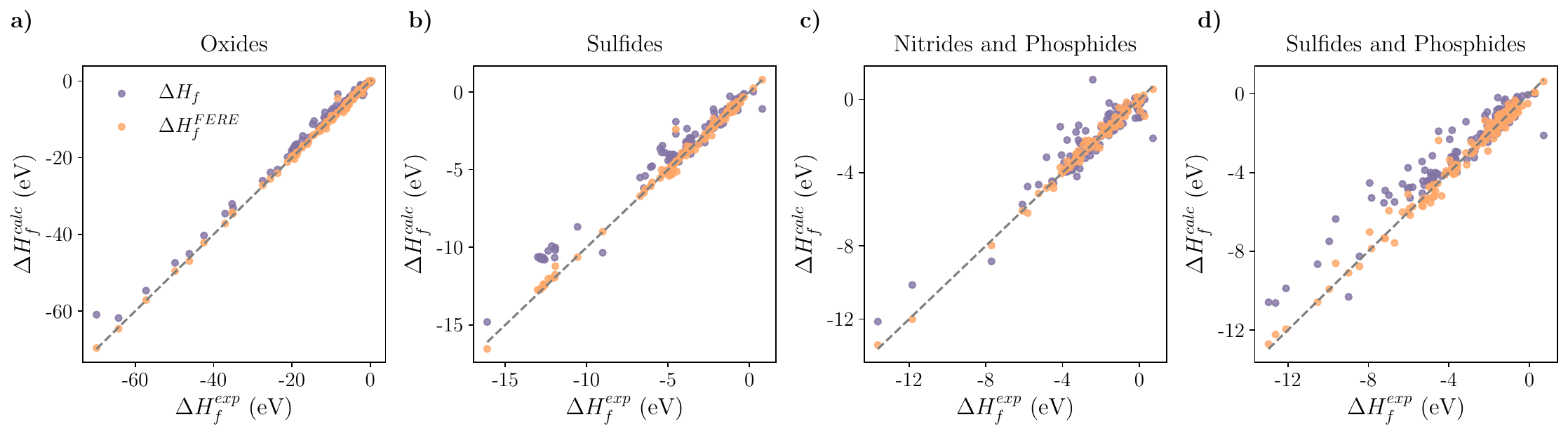}
    \caption{Values of $\Delta H_f^{calc}$ obtained with the FERE method compared to uncorrected values for the four considered datasets. A calculation of $\Delta H_f^{calc}$ based on Eq. \ref{eq:Hf} without fitted elemental corrections (purple data points) leads to a significant overestimation of $\Delta H_f^{calc}$ (less negative) for the majority of the compounds in the O-, S- and P-S datasets. Agreement with experimental values $\Delta H_f^{exp}$ is significantly increased when including fitted reference-phase energies in the formation enthalpy calculation (orange data points).}
    \label{fig:Comparison_FERE}
\end{figure*}

\subsubsection{Errors and uncertainty}
Phase predominance diagrams for the V-O, Ta-N, Sn-S, and Cu-P binary systems are chosen to span different chemistries and to probe systems with a large number of experimentally known crystalline compounds relative to other binary systems in their respective family.
Experimental values of $S$(298~K) and $C_{p}$(298~K) for a reference dataset containing 389 binary oxides, nitrides, phosphides and binary, ternary and quaternary sulfides are reproduced with mean absolute errors (MAE) of 0.042~meV/(atom$\cdot$K) and 0.043~meV/(atom$\cdot$K), respectively (see Fig.~\ref{fig:Comparison_S_Cp}). It is important to note that the experimentally available entropy data includes all sources of entropy including configurational. The relatively low error in the entropy calculated via MLIP phonons implies that considering only vibrational entropy in nominally ordered, non-alloyed compounds is a reasonable assumption.

The accuracy of DFT-calculated formation enthalpies compared to experimentally measured values is improved by introducing a fitted correction to the elemental reference phase energies (FERE)~\cite{FERE_paper}. Separate corrections are fitted for the oxide (150 compounds) and sulfide (118 compounds) binary systems. Due to the limited availability of experimental formation enthalpies for nitrides and phosphides, one common set of corrections is fitted to the phosphide-nitride set (98 pnictide compounds). The fitting scheme and results are detailed in the Methods section. MAEs of the computed formation enthalpy versus experiment are 0.236~eV/atom (O), 0.204~eV/atom (S) and 0.160~eV/atom (N-P) After the FERE corrections, the MAEs are reduced to 0.069~eV/atom, 0.058~eV/atom and 0.055~eV/atom, respectively (see Table~\ref{tab:FERE_MAE_different_datasets} and Fig.~\ref{fig:Comparison_FERE}). These errors are in line with previous work using the FERE method or similar correction schemes~\cite{zhangEfficientFirstprinciplesPrediction2018, pandeyHeatsFormationSolids2015, lanySemiconductorThermochemistryDensity2008}. For reference, a common estimate  for the error of experimentally measured formation enthalpies is 0.040~eV/atom~\cite{popleQuantumChemicalModels1999}.

It follows that the error on the formation enthalpy generally dominates over the heat capacity and entropy errors. It is only at temperatures above 1000~K that they become comparable. Hence the overall accuracy of our calculated phase predominance diagrams is primarily limited by the accuracy of the formation enthalpy prediction, even though this prediction is DFT-based rather than MLIP-based. This is an important conclusion, as it demonstrates that MLIP phonons calculated with the MatterSim model are not the main accuracy-limiting step.

\subsubsection{Comparison with experiment} \label{sec:binaries_comparison_experiment}

\begin{figure*}[]
    \centering
    \includegraphics[width=\textwidth]{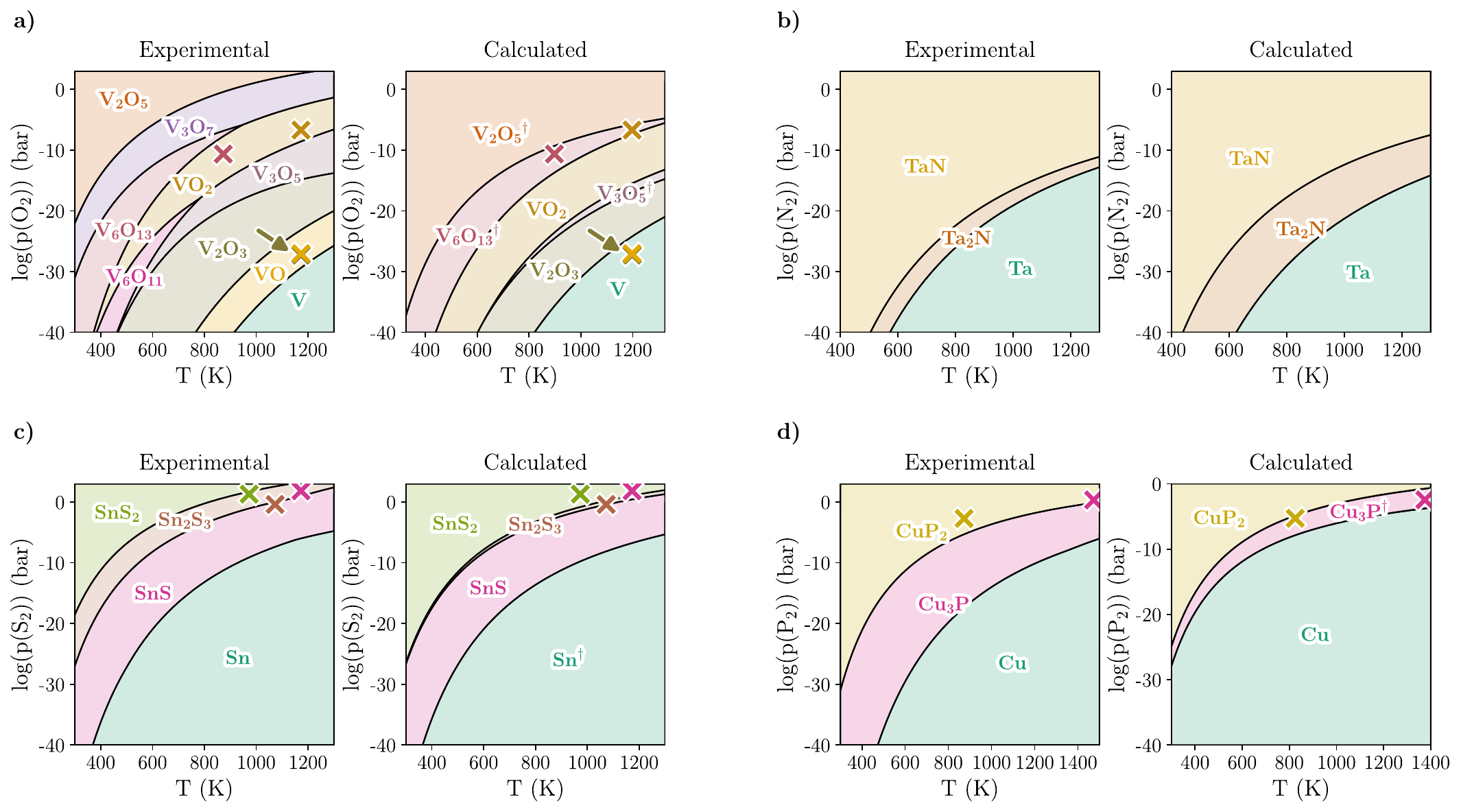}
    \caption{Exemplary phase predominance diagrams constructed for the V-O, Ta-N, Sn-S and Cu-P binary systems based on both experimentally measured and computationally derived thermochemical properties. Only binaries that are present in both the experimental and the computational database are considered here. Stability regions are color-coded by compound, with labels displayed in matching colors. All phases exhibiting imaginary frequencies according to MLIP phonon band structures are marked with a dagger ($^{\dagger}$), indicating potential dynamic instability~\cite{imaginary_phonon_frequencies}. Synthesis conditions for experimentally reported compounds are marked with a cross in a color representing the corresponding composition~\cite{burtonSynthesisCharacterizationElectronic2013, mootzKristallstrukturSn2S31967, odileCrystalGrowthCharacterization1978, nowotnyBeitragZurKenntnis1948, westmanNotePhaseTransition1961, aebiPhasenuntersuchungenImSystem1948}. Literature synthesis conditions for \ce{V2O3} and VO coincide and are thus marked with an additional arrow. We were unable to find bulk synthesis processes in the Ta-N system that employed \ce{N2} gas, hence the lack of crosses. The derivation of gas partial pressures from the reported synthesis conditions is detailed in the SI.}
\label{fig:comparison_experiment_oxides_nitrides_phosphides_sulfides}
\end{figure*}

Phase predominance diagrams for the V-O, Ta-N, Sn-S, and Cu-P binary systems are shown in Fig.~\ref{fig:comparison_experiment_oxides_nitrides_phosphides_sulfides} as a function of temperature and partial pressure of a relevant gaseous source for the respective anion (\ce{O2}, \ce{N2}, \ce{S2}, \ce{P2}). To assess the impact of the discussed uncertainties on the predicted predominance diagrams, phase boundaries derived using the workflow presented in this work are compared to those constructed solely from experimental thermochemical data in Fig.~\ref{fig:comparison_experiment_oxides_nitrides_phosphides_sulfides}. Only compounds included in both our chosen experimental thermochemical database (HSC Chemistry~\cite{HSCChemistry}) and our chosen computational database of DFT formation enthalpies (Ref.~\cite{PBEsol_DB} plus additional structures taken from either Materials Project~\cite{MP} or ICSD~\cite{zagoracRecentDevelopmentsInorganic2019} and relaxed by us) are considered.
In general, good agreement between the predicted and experimental phase boundaries is observed for all four binary systems. The majority of the experimentally synthesized phases are not only reproduced but also predicted stable under similar windows of gas partial pressures and temperatures in the experimental and computational thermochemical approaches.

It is interesting to see that temperature- and pressure-dependent thermodynamic modeling reveals stability windows for compounds that would be classified as metastable from a typical zero-temperature energy above hull approach. For example, two binary V-O compounds that have been experimentally synthesized (\ce{V3O5} and \ce{V6O11}) lie slightly above the zero-temperature stability hull ($E_{h}$ of 0.08~eV/atom and 0.05~eV/atom, respectively). Nevertheless, they are predicted to be stabilized over a broad range of temperatures by our computational approach.  
In the Cu-P binary system, \ce{Cu3P} is predicted to be thermodynamically stable at finite temperatures despite lying 27~meV/atom above the zero-temperature stability hull. This result explains why \ce{Cu3P} often forms more easily than \ce{CuP2}~\cite{crovettoCu3xPSemiconductorMetal2023} despite its zero-temperature metastability, as previously hypothesized~\cite{crovettoCrystallizeItIt2022}. These cases highlight the improved predictive capability of the presented approach for compound stability under realistic temperature and pressure conditions compared to zero-temperature stability metrics alone.

\begin{figure*}[]
    \centering
    \includegraphics[width=\textwidth]{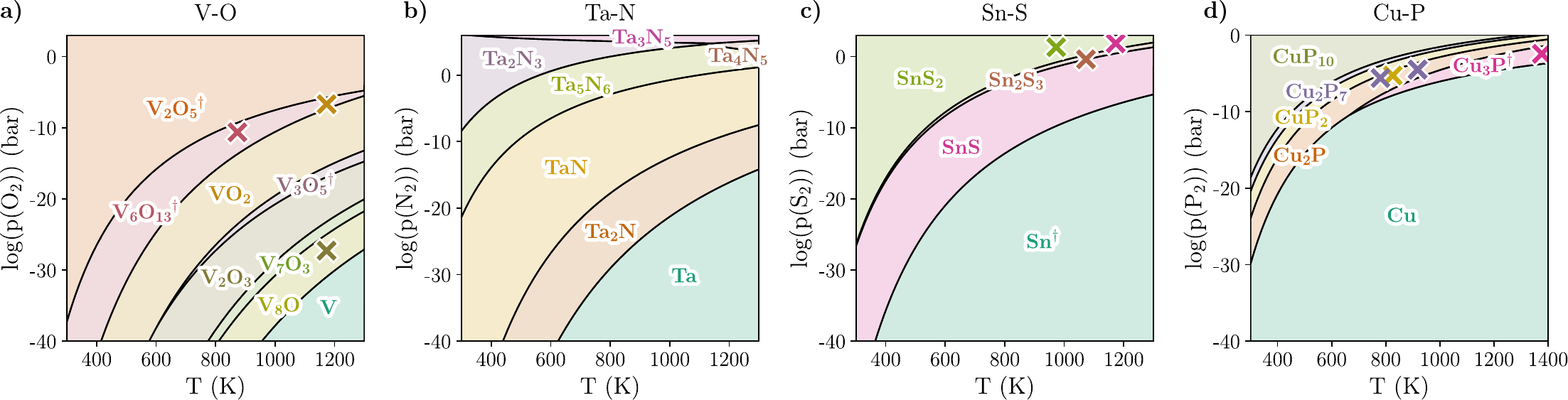}
    \caption{Phase predominance diagrams constructed for the V-O, Ta-N, Sn-S and Cu-P binary systems based solely on thermochemical properties calculated with our workflow. Here, all experimentally or computationally reported binary phases on Materials Project and the ICSD are considered, extending the diagrams to more materials than the ones with available experimental thermochemical data shown in Fig.~\ref{fig:comparison_experiment_oxides_nitrides_phosphides_sulfides}. Experimental synthesis conditions are marked with a cross~\cite{burtonSynthesisCharacterizationElectronic2013, mootzKristallstrukturSn2S31967, odileCrystalGrowthCharacterization1978, nowotnyBeitragZurKenntnis1948, westmanNotePhaseTransition1961, aebiPhasenuntersuchungenImSystem1948,moullerDarstellungEigenschaftenUnd1982}. See caption of Fig.~\ref{fig:comparison_experiment_oxides_nitrides_phosphides_sulfides} for further details.}\label{fig:oxides_nitrides_phosphides_sulfides_computational}
\end{figure*}

Uncertainties in the derived thermochemical quantities primarily lead to shifts in the locations of phase boundaries resulting in an increase or a reduction of the stability window size for given compounds as observed for, e.g., \ce{Ta2N} and \ce{Sn2S3} in in Fig.~\ref{fig:comparison_experiment_oxides_nitrides_phosphides_sulfides}. In some cases, an interplay of increase and reduction of stability window sizes for neighboring compounds can lead to individual phases disappearing from the predominance diagrams as it is the case for \ce{V3O7}, \ce{VO} and \ce{V6O11} in the V-O diagram.

For all phase boundaries included in Fig.~\ref{fig:comparison_experiment_oxides_nitrides_phosphides_sulfides}, the MAE in the temperature of calculated phase boundary positions versus experiment is 270~K. The Sn-S and Cu-P systems have the lowest and highest MAE, respectively (140~K and 520~K). The procedure used to calculate these errors is detailed in the SI. Experimentalists seeking to synthesize a compound can fix the gas partial pressure, choose a temperature in the middle of the predicted stability region of the targeted compound, and expect they they may need to vary the temperature by $\pm$270~K to hit the real synthesis window.

We hypothesize that the true error of the calculated phase boundary temperatures may be significantly lower than 270~K, because the experimental predominance diagrams in Fig.~\ref{fig:comparison_experiment_oxides_nitrides_phosphides_sulfides} are not based on direct observation of stability windows of different compounds. Instead, they are derived indirectly from experimental measurements of thermochemical properties of compounds. The ultimate test of the correctness of a predominance diagram is comparison with experimentally reported synthesis conditions, marked by crosses in Fig.~\ref{fig:comparison_experiment_oxides_nitrides_phosphides_sulfides}. Surprisingly, the agreement between stability windows and actual synthesis conditions is slightly better with our computationally determined phase predominance diagrams than with the experimental diagrams. Out of the eight compounds with experimental synthesis information in Fig.~\ref{fig:comparison_experiment_oxides_nitrides_phosphides_sulfides}, four fall within the computationally predicted stability windows (\ce{V6O13}, \ce{SnS2}, \ce{Sn2S3}, \ce{Cu3P}). Four additional compounds (\ce{VO2}, \ce{V2O3}, \ce{SnS} and \ce{CuP2}) are reasonably close, i.e., 50~K to 200~K off. This implies that the 270~K error for the phase boundary position quoted above is probably overestimated due to inaccuracies of the experimental benchmark.

\subsubsection{Predictive ability}

\begin{figure*}[]
    \centering
    \includegraphics[width=
    \textwidth]{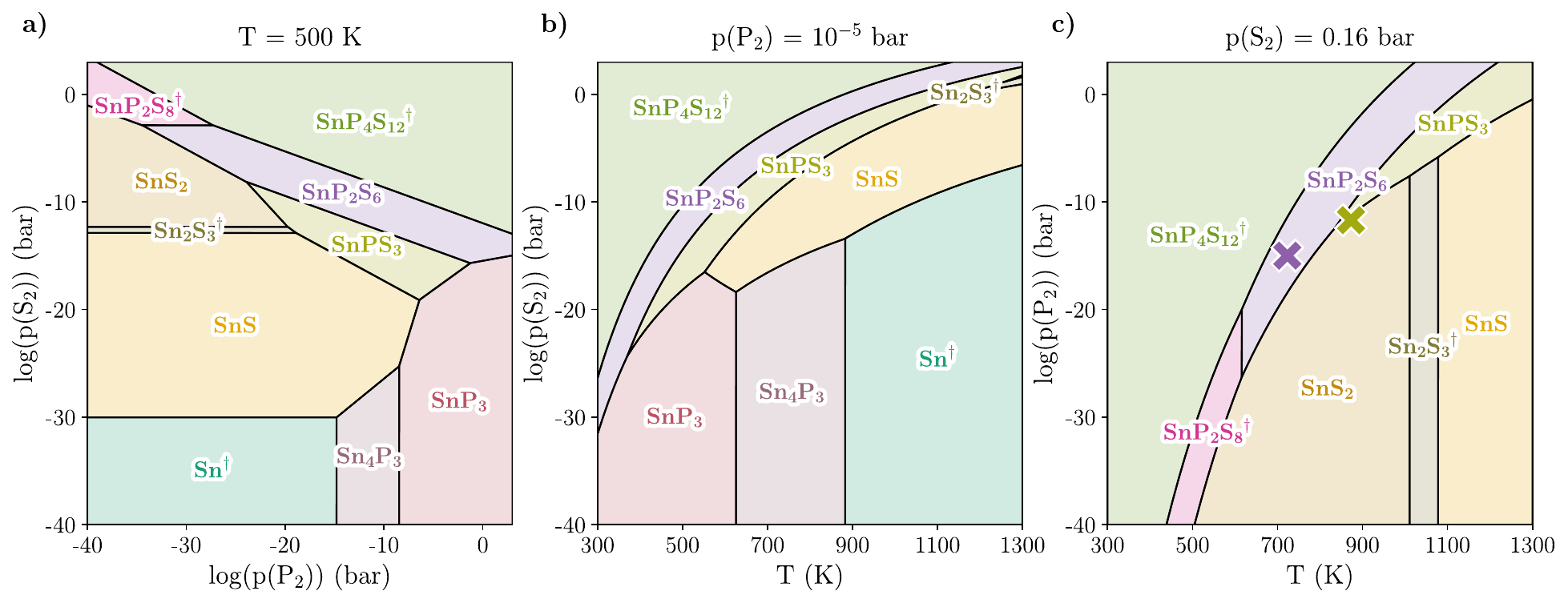}
    \caption{Exemplary representations of phase predominance diagrams for the Sn-P-S ternary system. a): Diagram for a constant temperature of 500~K in dependence of \ce{P2} and \ce{S2} partial pressures. b) and c): Diagrams dependent on temperature and \ce{S2} and \ce{P2} partial pressure, respectively, while the other is kept constant at \SI{e-5}{\bar} and 0.16~bar, respectively. Experimentally reported synthesis conditions of \ce{SnP2S6} and \ce{SnPS3}~\cite{heSnP2S6PromisingInfrared2022, dittmarStrukturDiZinnHexathiohypodiphosphatsSn2P2S61974} are marked with crosses in c). For the purpose of visualization, the fixed partial \ce{S2} pressure of 0.16~bar corresponds to the average of the \ce{S2} partial pressures used for the synthesis of \ce{SnP2S6} and \ce{SnPS3} (an order of magnitude from each other). This simplification does not affect the conclusions. The procedure used to extract synthesis conditions from the literature is detailed in the SI. See caption of Fig.~\ref{fig:comparison_experiment_oxides_nitrides_phosphides_sulfides} for further details.}
    \label{fig:phase_diagrams_example}
\end{figure*}

After assessing our method's limitations and its performance against experimental data, we now turn to using it as a predictive tool to identify synthesis conditions for new materials. We therefore construct a new set of predominance diagrams (Fig.~\ref{fig:oxides_nitrides_phosphides_sulfides_computational}) by considering all materials that are present either on Materials Project or the ICSD database and belong to the four binary systems studied here, even if they have no associated thermochemical data. This includes compounds that have not been synthesized before. The V-O system is expected to be particularly challenging, since V has a tendency for both mixed valence and magnetic ordering. Overall, we considered 28 V-O binaries and find finite stability windows for 7 of them (all true positives). \ce{V7O3} and \ce{V8O} are additions with respect to Fig.~\ref{fig:comparison_experiment_oxides_nitrides_phosphides_sulfides}, and we correctly predict finite stability windows for both. However, such windows occupy a region that was previously assigned to \ce{V2O3}, such that its experimental synthesis conditions are no longer reproduced in the purely computational diagram. In addition, there are 12 other V-O compositions that have been synthesized but do not appear in the predominance diagram (false negatives). We speculate that this is due to a combination of the complexity of V chemistry (see above) and the long history of synthesis in binary oxide systems, making it more likely that non-equilibrium routes have been found for the synthesis of thermodynamically unstable compounds. Under this assumption, one could argue that the earliest reported vanadium oxides are likely to be the ones with finite windows of thermodynamic stability, because they are more easily synthesized without the need to develop special reaction pathways for kinetic stabilization.

Indeed, the four V-O binaries present in the ICSD that were reported before 1950 (\ce{VO2}, \ce{V2O3}, \ce{V2O5} and \ce{V6O13}) all exhibit finite windows of thermodynamic stability in the computational predominance diagram in Fig. \ref{fig:oxides_nitrides_phosphides_sulfides_computational}. This indicates that the high occurrence of false negatives is not necessarily a flaw in the calculated diagrams, but may instead be a result of substantial efforts in the synthesis of metastable compounds in well-known materials systems.

For construction of the Ta-N predominance diagram in Fig.~\ref{fig:oxides_nitrides_phosphides_sulfides_computational}, 12 binaries are considered. Compared to Fig.~\ref{fig:comparison_experiment_oxides_nitrides_phosphides_sulfides}, the resulting diagram is extended by \ce{Ta2N3} and \ce{Ta5N6}, as well as \ce{Ta3N5} and \ce{Ta4N5} at particularly high \ce{N2} partial pressures. \ce{Ta2N}, \ce{TaN}, \ce{Ta3N5}, \ce{Ta4N5} and \ce{Ta5N6} have been synthesized, while \ce{Ta2N3} has not, giving 5 true positives and 1 potentially synthesizable new material. 2 additional Ta-N binaries have been experimentally reported and do not appear in the presented predominance diagram (2 false negatives). 

Although two additional theoretical phases (\ce{Sn3S} and \ce{Sn3S7}) are included in the construction of the Sn-S predominance diagram, neither of these two binaries is predicted to be stabilized under the conditions considered in this work. For the Sn-S system, all three phases present in the predominance diagrams have been experimentally reported. Nevertheless, two additional synthesized Sn-S binaries (\ce{Sn3S4} and \ce{Sn3S7}) do not appear, giving 3 true positives and 2 false negatives.

For the Cu-P binary system, a total of 10 binaries are considered, leading to significant differences with Fig.~\ref{fig:comparison_experiment_oxides_nitrides_phosphides_sulfides}. All four experimentally reported Cu-P binaries (\ce{CuP2}, \ce{CuP10}, \ce{Cu2P7} and \ce{Cu3P}) appear in the calculated diagram. A new phase \ce{Cu2P} is predicted to be stabilized across all temperatures, resulting in 4 true positives, zero false negatives, and 1 potentially synthesizable new material for this binary system. The inclusion of other binaries compared to Fig.~\ref{fig:comparison_experiment_oxides_nitrides_phosphides_sulfides} leads to a significant reduction in the \ce{CuP2} stability region. Nevertheless, the experimental synthesis conditions for \ce{Cu3P} are still accurately reproduced and the synthesis conditions for \ce{CuP2} lie reasonably close to the new phase boundary.

\subsection{Ternary systems with two gas-phase element sources: Phosphosulfides}

\subsubsection{High-throughput predominance diagrams}

\begin{figure*}[ht]
    \centering
    \includegraphics[width=\textwidth]{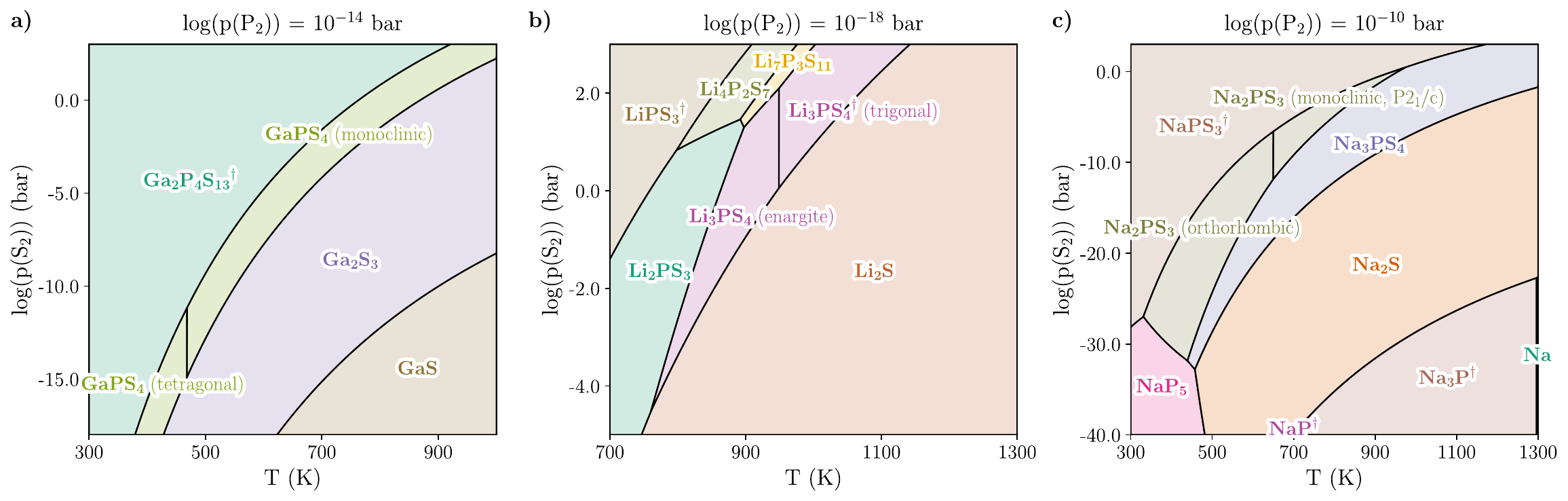}
    \caption{Calculated phase predominance diagrams for the Ga-P-S, Li-P-S, and Na-P-S phosphosulfide systems, predicting phase transitions between different polymorphs of the same composition (\ce{GaPS4}, \ce{Li3PS4}, and \ce{Na2PS3}).}
    \label{fig:example_different_stable_prototypes}
\end{figure*}

As a second case study, we show that our proposed method can readily be scaled to more than 1000 materials, even in the more complex case of multinary systems where two elements are supplied from the gas phase. The chemical space of interest is ternary metal phosphosulfides including (almost) all metals from the s, p and d-block.
In this space, approximately 60\% of the experimentally reported compounds lie on the zero-temperature convex hull ($E_h = 0$)~\cite{Javier_goat}. This suggests that zero-temperature thermodynamic stability is a more useful descriptor of synthesizability in phosphosulfides than in, e.g., oxides and nitrides~\cite{thermodynamic_limit}. Nevertheless, a number of experimentally realized compounds, such as \ce{CrPS4} ($E_h=0.03$~eV/atom)~\cite{diehlCrystalStructureChromium1977} and \ce{Cu7PS6} ($E_h = 0.05$~eV/atom)~\cite{kuhsArgyroditesNewFamily1979}
lie above the zero-temperature stability hull and would therefore be classified as metastable within a 0~K framework.

In the SI, we show predominance diagrams for 48 distinct metal phosphosulfide ternary systems, comprising over 1000 ternary compounds and about 450 binaries. A substantial fraction of these materials are hypothetical unsynthesized compounds, 19 of which lie on the 0~K convex hull~\cite{Javier_goat}. Similar to the binary systems, the accuracy of DFT-calculated formation enthalpies compared to experimentally measured values is improved by employing the FERE method. Since ternary phosphosulfides are absent from our experimental thermochemical database (HSC Chemistry), we apply the FERE method to 144 materials that are either phosphides (39) or sulfides (107), most of them binaries. With this approach, the MAE of the calculated formation enthalpy compared to experimentally measured values is reduced from 0.158~eV/atom to 0.072~eV/atom (see Table~\ref{tab:FERE_MAE_different_datasets} and Fig.~\ref{fig:Comparison_FERE}). Since the errors of MLIP calculations of entropy and heat capacity are about 0.04~meV/(atom$\cdot$K) (see above), using MLIP rather than DFT to calculate these quantities is again not the dominant source of inaccuracy at temperatures below $\sim$\SI{1500}{K}.

The average time needed for the high-throughput calculations is around 30~core-minutes per material for the DFT structure relaxation and additional 5~core-minutes per material for the MLIP phonon band structure and DOS. This time includes calculation of the derived thermochemical properties. Thus, MLIP determination of vibrational entropy and heat capacity is neither accuracy-limiting nor time-limiting for our calculation of phase predominance diagrams.

\subsubsection{The Sn-P-S ternary system as a case study}

A variety of representations of the phase predominance diagrams are possible for ternary phosphosulfides, owing to the multivariate dependence of the Gibbs free energy on temperature and partial pressures of the two assumed reactant gases (\ce{P2} and \ce{S2}). Fig.~\ref{fig:phase_diagrams_example} a) shows phase predominance in the Sn-P-S system as a function of the partial pressures of \ce{P2} and \ce{S2} at a fixed temperature of 500~K. As an alternative representation, Fig.~\ref{fig:phase_diagrams_example} b) and c) display phase predominance as a function of temperature and the partial pressure of either \ce{P2} or \ce{S2}, while the partial pressure of the other gas is held constant at \SI{e-5}{\bar} and 0.16~bar, respectively.

Binary Sn–S and Sn–P compounds appear in all displayed predominance diagrams. Additionally, the ternary compounds \ce{SnP2S8}, \ce{SnP4S12}, \ce{SnP2S6}, and \ce{SnPS3} are predicted to be the most stable phases under specific combinations of \ce{P2} and \ce{S2} partial pressures and temperatures. Of these ternary phosphosulfides, only \ce{SnPS3} and \ce{SnP2S6} have been experimentally reported to date~\cite{scottHighTemperatureCrystal1992, wangSynthesisCrystalStructure1995}. A comparison between literature derived experimental synthesis conditions for these compounds~\cite{heSnP2S6PromisingInfrared2022, dittmarStrukturDiZinnHexathiohypodiphosphatsSn2P2S61974} and the predicted synthesizability windows shown in Fig.~\ref{fig:phase_diagrams_example} c) gives remarkable agreement. While \ce{SnP2S6} and \ce{SnPS3} have DFT formation enthalpies that lie on the stability hull ($E_h = 0$) of the Sn–P–S system, \ce{SnP2S8} and \ce{SnP4S12} are located approximately 50~meV/atom above the hull and are thus considered metastable at 0~K and expected to decompose. In contrast, \ce{SnP2S7}, which lies closer to the stability hull ($E_h \approx 32$~meV/atom), does not appear in the phase predominance diagrams shown in Fig.~\ref{fig:phase_diagrams_example} and also exhibits imaginary phonon frequencies, indicating potential dynamic instability. These results emphasize again the importance of incorporating temperature-dependent thermodynamic properties when evaluating the experimental accessibility of predicted compounds.

\subsubsection{Predicting phase transitions}

The calculated predominance diagrams sometimes predict structural phase transitions between different polymorphs with the same composition. They arise from differences in the temperature dependence of thermochemical properties in the polymorphs, leading to reordering of their Gibbs free energies as temperature is varied. For our ternary phosphosulfide dataset, phase transitions are predicted for 55 compounds. For 19 of them, the transition occurs between 0~K and at room temperature. For the remaining 36, the transition occurs above room temperature. In general, the low-temperature polymorphs are often the ones that lie on the 0~K convex hull (lowest formation enthalpy structures). The high-temperature polymorphs are usually the ones with a higher entropic contribution to the Gibbs free energy.

Fig.~\ref{fig:example_different_stable_prototypes} shows exemplary predominance diagrams where phase transitions are predicted. Since the pressure dependent contribution to the Gibbs free energy of reaction of each compound depends solely on the stoichiometry (see Eq.~\ref{eq:pressure_contribution}), the temperature of the phase transition is not pressure dependent.
In Fig.~\ref{fig:example_different_stable_prototypes} a) the Ga-P-S system is displayed. At lower temperatures, tetragonal \ce{GaPS4}, which lies on the stability hull at 0~K, is the most stable prototype. At around 450~K, tetragonal \ce{GaPS4} is predicted to turn into a monoclinic structure ($E_h$ = 6~meV/atom at 0~K). This agrees well with reports of monoclinic \ce{GaPS4} synthesis at temperatures of 873-923~K~\cite{buckCrystalStructureGallium1973}.

Fig.~\ref{fig:example_different_stable_prototypes} b) shows the Li-P-S system. Below 950~K, the enargite phase of \ce{Li3PS4}, which lies on the 0~K stability hull and has been experimentally reported as a low temperature phase~\cite{hommaCrystalStructurePhase2011}, is predicted to be most stable. For higher temperatures, trigonal \ce{Li3PS4} ($E_h$ = 29~meV/atom at 0~K) is predicted to be stabilized. The trigonal structural prototype has not been experimentally reported for \ce{Li3PS4}. Nonetheless, phase transitions have indeed been observed for \ce{Li3PS4}. A highly disordered $\alpha-$\ce{Li3PS4} phase (not considered in this study due to disorder) has been reported for temperatures above 746~K~\cite{hommaCrystalStructurePhase2011, kaupImpactLiSubstructure2020}.

In Fig.~\ref{fig:example_different_stable_prototypes} c), the Na-P-S system is shown. Orthorhombic \ce{Na2PS3}, located on the zero-temperature stability hull, is predicted to be the most stable prototype in the low temperature regime. At higher temperatures \ce{Na2PS3} is expected to turn into a monoclinic structure ($E_h = $ 15~meV/atom, space group P2$_1$/c). Although this high-temperature monoclinic prototype has not been experimentally reported, a phase transition in \ce{Na2PS3} at elevated temperatures (430~K) has been observed~\cite{scholzPhaseFormationSynthetic2021}.
The experimentally reported low-temperature structure is within 2~meV/atom of the predicted low-temperature orthorhombic prototype, differing primarily by a reduction in crystallographic symmetry. The experimentally observed high-temperature polymorph of \ce{Na2PS3}~~\cite{scholzPhaseFormationSynthetic2021} is substantially different from the predicted one, and is characterized by substantial Na disorder. Thus, it was not considered in this study. 

\subsubsection{Trends in metal reactivity towards sulfur and phosphorus}

Interesting chemical trends can be extracted from a simple high-level analysis of the 48 predominance diagrams for ternary phosphosulfides shown in Figs.~\ref{fig:ternary_diagrams_1} to \ref{fig:ternary_diagrams_16}. Highly reactive metals with low electronegativity (alkalis, alkaline earths, Groups 3-4, and Al) are not thermodynamically stable as pure metals even at very low \ce{S2} and \ce{P2} partial pressures (below \SI{e-40}{bar} in many cases). This confirms that unwanted reactions of these metals with gas-phase elements may often be unavoidable in practice. Compatibly with their expected preference for ionic bonds, our predominance diagrams always predict that such metals will form sulfides rather than phosphides at equal \ce{S2} and \ce{P2} partial pressures. At the other end of the reactivity scale, the elemental phases of noble metals (e.g., Pt, Ir, Os, and especially Au) generally cover a large range of stability. Interestingly, less noble elements such as Bi, Sb, Ge, and Hg also have relatively wide stability windows as elemental species in \ce{S2} and \ce{P2} atmospheres.
The only elements that are more prone to form phosphides than sulfides are certain transition metals (Cu, Co, Mn, Ni, Pd).

Phosphosulfide materials with equal fractions of S and P (e.g., LaPS, CoPS) tend to be stabilized by partial pressures of \ce{S2} and \ce{P2} that are similar to each other, as expected. However, if the metal is very electropositive (e.g., Y, La, Zr) this window of stability shifts to higher \ce{P2} pressures due to the high stability of binary sulfides. For all these MPS compounds, reactive gas partial pressures should neither be too low nor too high. In the first case, binaries are stabilized. In the second case, thiophosphate compounds with much higher S/P atomic ratios are stabilized. As expected, thiophosphate compounds tend to form stability windows that are more on the high \ce{S2} partial pressure side, due to their high sulfur content.
Finally, we note that the stability windows of novel phosphosulfide compositions such as CuPS and \ce{Ca4P2S} are often very narrow, requiring precise experimental control and further demonstrating the practical applicability of calculated predominance diagrams.

\subsubsection{Computational discovery of materials and their synthesizability windows}

\begin{figure}[]
    \centering
    \includegraphics[width=0.8\columnwidth]{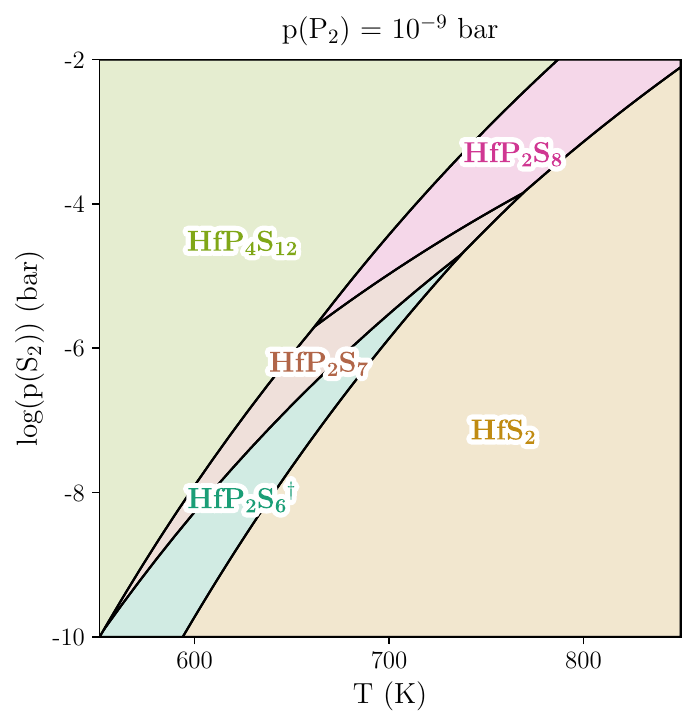}
    \caption{Phase predominance diagram for the Hf-P-S ternary system showing temperature and partial pressure conditions under which the computationally discovered \ce{HfP4S12}, \ce{HfP2S7}, \ce{HfP2S8} and the experimentally reported \ce{HfP2S6} structures are predicted to be stable.}
    \label{fig:Hf_phase_diagram}
\end{figure}

Discovery of new materials and quantitative prediction of their synthesizability windows can be conducted in a single workflow. As an example, we identified 19 previously unknown phosphosulfide materials predicted to be on the zero-temperature convex hull~\cite{Javier_goat},. Among them, four belong to the Hf-P-S system (\ce{HfP4S12}, \ce{HfP2S7}, \ce{HfP2S8}, and \ce{Hf2P3S}), making this system of particular interest for the synthesis of novel compounds. Only \ce{HfP2S6} had been experimentally and computationally reported in this material system prior to our work~\cite{simonDarstellungUndAufbau1985}. \ce{HfP2S6} lies 5~meV/atom above the stability hull. Fig.~\ref{fig:Hf_phase_diagram} shows a phase predominance diagram calculated for the Hf–P–S ternary system including these four new compounds. Three out of four are also predicted to be thermodynamically stable at finite temperatures over a range of partial pressures of the reactant gases. In addition, \ce{HfP2S6} is stabilized over a finite temperature window and a range of \ce{S2} partial pressures in spite of its metastability at 0~K, consistent with its experimental observation.

\section{Conclusion and Outlook}
We demonstrated an accelerated workflow to generate computational phase predominance diagrams for any inorganic material system. These diagrams are useful quantitative estimations of synthesizability windows of materials in a parameter space that is easily understood by experimentalists (temperature versus gas partial pressures).
The key step enabling high-throughput predominance diagram generation was the determination of vibrational entropy and heat capacity of materials from phonon band structures calculated via machine learned interatomic potentials (MLIP). Despite the semi-empirical nature of MLIP methods, the present results indicate that they do not compromise either the accuracy or the computational time needed to produce phase predominance diagrams based on DFT-calculated formation enthalpies.

Predominance diagrams generated in oxide, nitride, sulfide, and phosphide binary systems were generally consistent with experimental thermochemical data and experimentally reported synthesis conditions, highlighting chemical versatility. As expected, compounds that are metastable at 0~K are sometimes found to be thermodynamically stable under finite-temperature conditions, e.g., \ce{VO}, \ce{V3O5}, \ce{V6O11}, and \ce{Cu3P}. In a similar fashion, materials lying on the convex stability hull at 0~K may no longer be stable at elevated temperatures.

As a high-throughput application of our method, we generated predominance diagrams for 48 distinct metal phosphosulfide ternary systems, comprising over 1000 ternary compounds and about 450 binaries. Some highlights were the identification of thermodynamically driven phase transitions between polymorphs, broad trends in metal reactivity with sulfur and phosphorus, and the quantitative prediction of synthesis conditions for computationally discovered materials. These features are expected to facilitate communication between theorists and experimentalists, and between the computational and synthetic tasks of closed-loop autonomous labs for materials discovery. Future work could focus on the incorporation of configurational entropy, energy-lowering defects, and kinetic effects.

\section{Methods}\label{section:methods}

\subsection{Data provenance}
The phase predominance diagram of a generic binary or ternary system shows the solid phase with the lowest Gibbs free energy $\Delta G(T, p_i, p_j...)$ among all considered phases in that system at each temperature $T$ and gas partial pressure $(p_i, p_j,...)$ coordinate point.
The phases that are considered in the four chosen binary systems are the union of the material sets present on Materials Project~\cite{MP}, the Inorganic Crystal Structure Database (ICSD)~\cite{zagoracRecentDevelopmentsInorganic2019}, and a computational DFT database at the PBEsol level~\cite{PBEsol_DB}. A full tabulated list of materials is given in the SI. The PBEsol database is also used for the crystal structures and enthalpies of elemental reference phases. For consistency, we relaxed all the considered entries from Materials Project and ICSD using the PBEsol functional.
Potential ternary phosphosulfide (M-P-S) phases are taken from our own open-access computational database at the PBEsol level, containing about 1000 entries from a prototype-based screening study.~\cite{Javier_goat}

The Gibbs free energy $\Delta G(T, p_i, p_j,...)$ for each considered material is calculated with the HSC Chemistry software package (Metso)~\cite{HSCChemistry}. Three thermochemical properties are required for each material: its entropy $S$(298~K) and formation enthalpy $\Delta H_f$(298~K) at room-temperature, and its temperature-dependent heat capacity $C_p(T)$. Calculation details for these three quantities are given in the next sections, with more details in the SI.

\subsection{Formation enthalpy by density functional theory with reference energy correction} \label{subsec:theoretical_dataset} \label{subsec:FERE}

All structures that do not originate from the general-purpose PBEsol database~\cite{PBEsol_DB} or from our own PBEsol database of ternary phosphosulfides~\cite{Javier_goat} are relaxed using DFT. We employ the projector-augmented wave method~\cite{blochlProjectorAugmentedwaveMethod1994} and the PBEsol functional~\cite{perdewRestoringDensityGradientExpansion2008} as implemented in the Vienna Ab-Initio Simulation Package (VASP)~\cite{kresseInitioMolecularDynamics1993, kresseInitioMoleculardynamicsSimulation1994, kresseEfficiencyAbinitioTotal1996, kresseEfficientIterativeSchemes1996, kresseUltrasoftPseudopotentialsProjector1999}. All structure relaxations are performed using a $\Gamma$-centered uniform k-point grid with 5 k-points per Å$^{-1}$ along each reciprocal lattice direction by setting the KSPACING tag in VASP to 0.22. An energy cutoff of at least 550~eV is selected. As convergence criteria, a difference of 10~meV/Å force on all atoms, and a difference of 10$^{-6}$~eV/atom in total energy are chosen~\cite{Javier_goat}. 

The enthalpy of formation of a given compound is defined as the difference between the compound's zero-temperature total energy and the energies of its elemental components $\mu_i$, weighted by their stoichiometric coefficients $n_i$.
\begin{equation}
    \Delta H_f = E_{Total} - \sum_i n_i\mu_i \quad .
    \label{eq:Hf}
\end{equation}
Even though a material's zero-temperature formation enthalpy is directly available from a simple DFT structural relaxation of the material and of its elemental constituent phases, reproducing experimental formation enthalpies at room temperature is a known challenge~\cite{enthalpy_of_formation_1, wangFrameworkQuantifyingUncertainty2021}.
A convenient approach to improve the formation enthalpy prediction is using fitted elemental-phase reference energies (FERE)~\cite{FERE_paper}. In this method, corrections $\delta\mu_i^{FERE}$ are fitted to the total energies of each of the elemental reference phases to reproduce experimentally observed values of $\Delta H_f$ 
\begin{equation}
    \mu_i^{FERE} = \mu_i + \delta \mu_i^{FERE} \quad .
\end{equation}
Values for these corrections are obtained by fitting the $\delta \mu_i^{FERE}$ correction for each element to minimize the difference between experimental and calculated $\Delta H_f$ ($\Delta H_f^{exp} - \Delta H_f^{calc}$) for all materials with available experimental formation enthalpies. The corrections for all elements are fitted simultaneously
\begin{equation}
    \Delta H_f^{exp} - \Delta H_f^{calc} = \sum_i n_i \delta \mu_i^{FERE} \quad .
    \label{linear least squares}
\end{equation}
This requires a reference dataset containing both experimental values for the formation enthalpy and the corresponding DFT total energies for each of the compounds, raising the question of how to correctly compare experimental values (typically available at room temperature) with DFT value (at 0~K). It has been shown~\cite{enthalpy_of_formation_1} that including the zero-point energy and enthalpic contributions at finite temperatures in the calculation does not lead to a significant improvement of the formation enthalpy prediction. Thus, we simply use the DFT-calculated total energy at 0~K in the fitting of the elemental reference-phase energies.

Experimentally measured values for the formation enthalpy $\Delta H_f^{exp}$ are extracted from the HSC Chemistry experimental database, which provides experimental $H$(298~K), $S$(298~K) and $C_p$ data for about 30,000 species~\cite{HSCChemistry}. Four distinct sets of elemental reference phase energies are fitted separately using the FERE approach. The fitted energies are shown in Table~\ref{tab:elemental_corrections}. The first set consists of 150 binary metal oxides, and it is used to plot the V-O predominance diagram. the second set consists of 118 binary metal sulfides, used to plot the Sn-S predominance diagram. The third set includes  binary metal phosphides (46) and nitrides (53), used to plot the Ta-N and Cu-P predominance diagrams. The reason for fitting nitrides and phosphides together is the scarcity of experimental thermochemical data for these classes of materials. This choice may be further justified by homovalence of N and P. The fourth dataset consists of metal phosphides (39) and metal sulfides (107), and includes 121 binaries, 21 ternaries, and 2 quaternaries. It is used to plot all ternary phosphosulfide predominance diagrams.

\subsection{Entropy and heat capacity by machine-learned interatomic potentials} \label{subsec:MatterSim}

Room-temperature vibrational entropy $S$(298~K) and heat capacity $C_p$ for each material are calculated employing the universal MLIP model MatterSim~\cite{MatterSim} combined with the phonopy package~\cite{phonopy_1} using 20$\times$20$\times$20 Monkhorst-Pack k-point meshes.
4$\times$4$\times$4 supercells with atoms displaced by $0.01$~Å are generated using phonopy and the resulting forces needed to calculate phonon frequencies and dispersion curves are predicted using MatterSim \cite{MatterSim}. The temperature-dependent entropy $S(T)$ and constant-volume heat capacity $C_V(T)$ are derived from the Helmholtz free energy and harmonic phonon energy, respectively. These quantities are derived via summation over all calculated phonon modes. Although the quantity derived from the phonon band structure is the constant-volume heat capacity $C_V$, we assume $C_V \simeq C_p$ under the assumption of incompressible solids~\cite{Ba_Zr_S_phase_diagrams}.

The room-temperature entropy $S$(298~K) and temperature dependent heat capacity $C_p(T)$ from 298~K to 1300~K are computed by phonopy in steps of 25~K for all considered DFT-relaxed crystal structures. 
In HSC Chemistry, the temperature dependence of the heat capacity is described by the extended Kelley equation with fitted parameters A-F \cite{extended_kelley_equation}. 
\begin{equation}
\begin{aligned}
   C_p(T) = & A + BT \cdot 10^{-3} + \frac{C}{T^2}\cdot 10^{5} + D T^2 \cdot 10^{-6} \\
   & + \frac{E}{T^3}\cdot 10^{8} + FT^3\cdot 10^{-9} \quad .
   \label{eq:kelley}
\end{aligned}
\end{equation}
Therefore, the temperature-dependent $C_p(T)$ obtained from the phonon calculation is fitted with the Kelley equation in the 298~K $\rightarrow$ 1300~K range, and the resulting best-fit parameters A-F are then supplied to HSC Chemistry to calculate Gibbs free energies (see next section).

The room-temperature heat capacity and entropy values calculated for all binary compounds in the four datasets utilized in the fitting of the elemental reference-phase energy corrections are compared to the corresponding experimental values available in HSC Chemistry (389 binaries for entropy, 346 binaries for heat capacity). The MAEs are 0.042~meV/(atom$\cdot$K) for $S$(298~K) and 0.043~meV/(atom$\cdot$K) for $C_p$(298~K). 

\subsection{Calculation of Gibbs free energy and phase predominance diagrams by classical thermochemistry}

Two-dimensional phase predominance diagrams are constructed in HSC Chemistry after extracting the temperature- and partial pressure-dependent Gibbs free energy $\Delta G(T, p_i, p_j,...)$ from the formation enthalpy and entropy at a single temperature, and from the temperature-dependent heat capacity ($\Delta H_f$(298K), $S$(298K), and $C_p(T)$, respectively). Experimental predominance diagrams are based on the experimental values of these three quantities for the phases in the V-O, Ta-N, Sn-S, and Cu-P systems for which thermochemical data is available in HSC Chemistry. Computational diagrams are based on the DFT-calculated $\Delta H_f$ after applying the elemental reference phase correction, and on the MLIP-calculated $S$(298K), and $C_p(T)$.
Assuming solids are incompressible, the dependence of the Gibbs free energy on the total pressure of the system can be neglected~\cite{Ba_Zr_S_phase_diagrams}. Hence, the temperature dependent Gibbs energy of a given compound is defined by
\begin{equation}
    G(T) = H(T) - T\cdot S(T)
    \label{eq:gibbs}
\end{equation}

The temperature-dependent enthalpy $H(T)$ can be calculated from the formation enthalpy at a single temperature and the temperature-dependent heat capacity as

\begin{equation}
    H(T) = \Delta H_f (298 K) + \int_{298K}^T C_p(T) dT 
\end{equation}

The temperature-dependent entropy $S(T)$ can be derived with a similar procedure as
\begin{equation}
    S(T) = S(298K) + \int_{298K}^T \frac{C_p(T)}{T}dT 
\end{equation}
The Gibbs energy of a compound ($G_c$) is then referenced to the Gibbs free energy of its elemental reference phases ($G_{i,r}$) as
\begin{equation}
    \Delta G^0 = n_{c} G_{c} - \sum_i n_i G_{i, r} \quad .
\end{equation}
with $n_c$ the number of atoms in the compound and $n_i$ the stoichiometric coefficient of the elemental constituents. The Gibbs free energy of each elemental phase $G_i$ is calculated with Eq.~\ref{eq:gibbs}, using the experimental $\Delta H_f$(298K), $S$(298K), and $C_p(T)$ values available in HSC Chemistry.

Finally, the Gibbs free energy dependence on the partial pressure of reactive gases is defined by the equilibrium constant K
\begin{equation}
    K = \frac{a_{c}^{n_{c}}}{\prod_i a_{i, r}^{n_{i}}}
    \label{eq:equilibrium_1}
\end{equation}
with $a$ being the thermodynamic activity of the corresponding species and $n$ the stoichiometric coefficient. For gaseous substances, $a$ is described by their effective partial pressure $p$ and for pure substances in the solid state, $a$ is equal to 1~\cite{millsQuantitiesUnitsSymbols1993, ewingIupacReportIUPAC1995}. In this work, phase predominance diagrams are calculated depending on one or two partial pressures of the corresponding reactive gases (\ce{O2}, \ce{N2}, \ce{P2}, \ce{S2}) and the metal activity is assumed to be 1 over the total temperature range.
This implies that sublimation of metallic species during phase formation is neglected and may lead to an overestimation of the thermodynamic stability of metal-rich compounds at elevated temperatures, particularly for systems containing highly volatile metals.

The equilibrium constant K is connected to the Gibbs free energy by
\begin{equation}
    \ln{K} = -\frac{\Delta G_R}{RT} \quad .
    \label{eq:equilibrium_2}
\end{equation}
Hence, the total Gibbs free energy for a compound is extracted by combining Eq.~\ref{eq:gibbs} and Eq.~\ref{eq:equilibrium_2} ~\cite{HSCChemistry, epifanoEllinghamDiagramNew2023}
\begin{equation}
    \Delta G = \Delta G^0 + RT\ln{K} \quad .
    \label{eq:pressure_contribution}
\end{equation}
The phase predominance diagrams are then constructed so that the compound with the minimal $\Delta G$ at a given temperature and partial pressure coordinate point is displayed.

\begin{acknowledgments}
This work was co-funded by the European Union (ERC, IDOL, 101040153). Views and opinions expressed are however those of the authors only and do not necessarily reflect those of the European Union or the European Research Council. Neither the European Union nor the granting authority can be held responsible for them. This work was supported in part by a research grant (42140) from VILLUM FONDEN.
\end{acknowledgments}
\vspace{0.5cm}
\section*{Data availability}
The computational data generated from this study is available in the NOMAD database at \url{https://doi.org/10.17172/nomad.1czd-xs7s}. It includes jupyter notebooks that employ the FERE method for the four presented datasets and enable visualization of phase predominance diagrams for ternary phosphsulfides, along with the raw phase predominance diagram data. Additionally, an excel sheet with all experimental synthesis conditions considered in this work, as well as the full database of derived thermodynamic properties ($\Delta H_f$, $S(298)$ and Kelley coefficients $A$ - $F$ describing $C_p(T)$) for all compounds and elemental phases considered in the construction of phosphosulfide phase predominance diagrams, are provided.

\FloatBarrier
\bibliography{bib.bib}

\clearpage
\newpage


\onecolumngrid
\begin{center}
\begin{large}
\textbf{SUPPORTING INFORMATION \\}
\vspace{0.5cm}
\textbf{Rapid estimation of synthesizability windows of inorganic materials from first principles\\}
\end{large}
\vspace{0.5cm}
Finja Tadge, Javier Sanz Rodrigo, Andrea Crovetto
\end{center}

\vspace{1cm}

\setcounter{page}{1}
\renewcommand*{\thepage}{S\arabic{page}}
\renewcommand{\thefigure}{S\arabic{figure}}
\renewcommand{\thetable}{S\arabic{table}}

\setcounter{figure}{0}
\setcounter{section}{0}
\setcounter{table}{0}

\section{Calculation of Thermochemical properties}

\subsection{FERE approach: Corrections to the elemental reference phase energies}

Agreement between DFT derived total energies and experimentally measured formation enthalpies is improved by employing the FERE approach~\cite{FERE_paper}. The PBE total energies $\mu_i$, along with the distinct elemental corrections $\delta \mu_i^j$ that are fit for the four different datasets are given in Table~\ref{tab:elemental_corrections}. $\delta \mu_i^{O}$ is the corection for the oxide dataset, $\delta \mu_i^{S}$ for the sulfide dataset, $\delta \mu_i^{N-P}$ for the nitride-phosphide dataset, and $\delta \mu_i^{P-S}$ for the phosphide-sulfide dataset used for the ternary phosphosulfide predominance diagrams.

\begin{table}[h!]
\centering
\begin{tabular}{c c c c c c | c c c c c c }
\hspace*{3mm}Element\hspace*{3mm} & \hspace*{3mm}$\mu_i$ \hspace*{3mm}& $\delta \mu_i^{O}$\hspace*{3mm} & \hspace*{3mm} $\delta \mu_i^{S}$ \hspace*{3mm} & \hspace*{3mm} $\delta \mu_i^{N-P}$\hspace*{3mm}  & \hspace*{3mm} $\delta \mu_i^{P-S}$ \hspace*{3mm} & \hspace*{3mm}Element\hspace*{3mm} & \hspace*{3mm}$\mu_i$ \hspace*{3mm}& \hspace*{3mm}$\delta \mu_i^{O}$\hspace*{3mm}& \hspace*{3mm}$\delta \mu_i^{S}$\hspace*{3mm} & \hspace*{3mm}$\delta \mu_i^{N-P}$\hspace*{3mm} & \hspace*{3mm}$\delta \mu_i^{P-S}$\hspace*{3mm} \\ 
\hline \hline
Ac & -4.493 & 1.192 & - & - & - & Nd & -5.167 & 0.873 & 0.448 & 0.266 & - \\
Ag & -3.311 & -0.086 & 0.050 & -0.067 & 0.146 & Ni & -6.16 & 1.476 & -0.165 & -0.041 & -0.157 \\
Al & -4.092 & 0.929 & 0.410 & 0.448 & 0.785 & O & -5.249 & 0.037 & 0.058 & - & - \\
As & -5.106 & 1.075 & -0.041 & - & - & Os & -12.216 & -0.638 & -0.274 & -0.250 & -0.386 \\
Au & -3.924 & -0.243 & -0.177 & 0.174 & -0.099 & P & -5.791 & - & -  & 0.043 & 0.122 \\
B & -7.016 & 0.464 & - & -0.454 & -0.480 & Pa & -10.231 & 0.584 & - & - & - \\
Ba & -2.149 & 0.785 & 0.468 & 0.452 & 0.254 & Pb & -3.999 & 0.123 & -0.185 & - & -0.230 \\
Be & -4.000 & 0.708 & 0.096 & 0.188 & - & Pd & -5.961 & 0.201 & -0.173 & - & -0.247 \\
Bi & -4.285 & -0.048 & -0.401 & - & -0.468 & Pm & -5.137 & 0.760 & 0.594 & - & - \\
Ca & -2.134 & 0.617 & 0.449 & 0.133 & 0.143 & Pr & -5.185 & 1.234 & 0.406 & 0.423 & - \\
Cd & -1.197 & 0.591 & 0.291 & 0.218 & -0.024 & Pt & -6.964 & -0.560 & -0.307 & - & -0.374 \\
Ce & -6.594 & 1.047 & 0.588 & 0.281 & - & Pu & -15.154 & 1.679 & 1.633 & 1.140 & - \\
Co & -7.737 & 0.782 & -0.265 & -0.012 & -0.289 & Rb & -1.025 & 0.180 & 0.198 & -0.705 & - \\
Cr & -10.242 & 0.942 & -0.061 & -0.064 & -0.001 & Re & -13.332 & -0.046 & -0.204 & - & -0.293 \\
Cs & -0.943 & 0.147 & 0.120 & -0.514 & - & Rh & -8.138 & -0.101 & -0.305 & - & -0.372 \\
Cu & -4.334 & 0.329 & 0.094 & 0.475 & 0.282 & Ru & -10.172 & -0.441 & -0.114 & - & -0.203 \\
Dy & -4.978 & 0.505 & 0.675 & - & - & S & -4.360 & 0.194 & 0.129 & 0.292 & 0.173 \\ 
Er & -4.927 & 0.567 & 0.751 & -0.118 & - & Sb & -4.557 & 0.369 & -0.170 & - & -0.174 \\ 
Eu & -10.548 & 1.182 & -0.288 & 0.540 & - & Sc & -6.584 & 0.795 & 0.839 & -0.858 & 0.777 \\
Fe & -8.963 & 1.072 & -0.464 & -0.128 & -0.237 & Se & -3.850 & 0.194 & - & - & - \\
Ga & -3.353 & 0.842 & 0.745 & 0.418 & 0.866 & Si & -5.747 & 0.975 & 0.084 & -0.151 & 0.079 \\
Gd & -14.438 & 0.652 & -0.022 & -0.032 & - & Sm & -5.092 & 0.773 & 0.708 & 0.012 & - \\
Ge & -4.940 & 0.317 & 0.008 & 0.881 & -0.175 & Sn & -4.260 & 0.555 & 0.202 & - & 0.116 \\
Hf & -10.503 & 0.711 & 0.975 & -0.036 & 0.886 & Sr & -1.840 & 0.604 & 0.340 & 0.407 & 0.296 \\
Hg & -0.567 & 0.326 & 0.133 & - & 0.089 & Ta & -12.551 & 0.553 & 0.021 & -0.124 & -0.090 \\
Ho & -4.950 & 0.540 & 0.695 & -0.253 & - & Tb & -5.010 & 1.178 & 0.628 & - & - \\
In & -2.984 & 0.758 & 0.233 & 0.254 & 0.289 & Tc & -11.259 & -0.132 & -0.110 & - & -0.199 \\
Ir & -9.822 & 0.079 & -0.268 & - & -0.339 & Te & -3.526 & 0.114 & - & - & - \\
K & -1.133 & 0.318 & 0.373 & -0.655 & 0.308 &  Th & -8.107 & 0.819 & 0.045 & 0.567 & - \\
La & -5.346 & 0.598 & 0.665 & 0.413 & 0.541 & Ti & -8.322 & 0.490 & -0.394 & -0.266 & -0.461 \\
Li & -1.968 & 0.344 & 0.184 & -0.033 & 0.161 & Tl & -2.630 & 0.141 & 0.108 & 3.812 & 0.032 \\
Lu & -4.872 & 0.349 & 0.878 & - & - & Tm & -4.899 & 0.504 & 0.806 & - & - \\
Mg & -1.717 & 0.968 & 0.475 & 0.529 & 0.471 & U & -12.158 & 0.764 & 0.866 & 0.488 & - \\
Mn & -9.708 & 1.141 & 0.483 & 0.032 & 0.069 & V & -9.634 & 0.240 & -0.234 & -0.457 & -0.279 \\
Mo & -11.768 & -0.002 & -0.184 & -0.185 & -0.255 & W & -13.807 & 0.232 & -0.312 & 0.277 & -0.401 \\
N & -8.525 & - & - & -0.105 & - & Y & -6.822 & 0.873 & 0.865 & -0.580 & 0.803 \\
Na & -1.400 & 0.433 & 0.243 & -0.671 & 0.221 & Zn & -1.582 & 0.700 & 0.330 & 0.184 & 0.158 \\
Nb & -10.821 & 0.830 & 0.016 & -0.167 & -0.051 & Zr & -9.090 & 0.854 & 0.784 & -0.048 & 0.713 \\
\hline
\hline
\end{tabular}
\caption{Elemental reference phase energies $\mu_i$ taken from the computational PBEsol database~\cite{PBEsol_DB}, along with the fitted corrections to the elemental reference phase energies $\delta \mu_i^{j}$ for four different datasets considered in this work. All energies are given in (eV/atom).}
\label{tab:elemental_corrections}
\end{table}

\begin{table}[h!]
\centering
\begin{tabular}{c c c | c c c}
\hspace*{3mm}Compound\hspace*{3mm} & \hspace*{12mm} Source \hspace*{12mm}& \hspace*{3mm} ID \hspace*{3mm} & \hspace*{3mm}Compound\hspace*{3mm} & \hspace*{12mm} Source \hspace*{12mm}& \hspace*{3mm} ID \hspace*{3mm}\\ 
\hline \hline  
VO$^*$ & PBEsol DB & - & TaN$^*$ & PBEsol DB & - \\
\ce{VO2}$^*$ & PBEsol DB & - & \ce{TaN2} & Materials Project & mp-1019272 \\
\ce{VO3}($^*$) & Materials Project & mp-1207792 & \ce{TaN4}$^*$ & ICSD & Collection Code 122049 \\
\ce{V2O3}$^*$ & PBEsol DB & - & \ce{TaN5}$^*$ & ICSD & Collection Code 122050\\
\ce{V2O5}$^*$ & PBEsol DB & - & \ce{Ta2N}$^*$ & PBEsol DB & - \\
\ce{V3O5}$^*$ & PBEsol DB & - & \ce{Ta2N3} & Materials Project & mp-1208406 \\ 
\ce{V3O8}$^*$ & Materials Project & mp-796383 & \ce{Ta3N} & Materials Project & mp-1217984 \\
\ce{V3O10}$^*$ & Materials Project & mp-818205 & \ce{Ta3N2} & PBEsol DB & - \\
\ce{V4O7}$^*$ & Materials Project & mp-555597 & \ce{Ta3N5}$^*$ & PBEsol DB & - \\
\ce{V4O9$^*$} & Materials Project & mp-27412 & \ce{Ta4N3} & Materials Project & mp-1218021 \\
\ce{V5O7} & Materials Project & mp-1046039 & \ce{Ta4N5}$^*$ & PBEsol DB & - \\
\ce{V5O9}$^*$ & Materials Project & mp-542334 & \ce{Ta5N6}$^*$ & Materials Project & mp-1642 \\
\ce{V5O12} & PBEsol DB & - & SnS$^*$ & PBEsol DB & - \\
\ce{V6O11}$^*$ & Materials Project & mp-30518 & \ce{SnS2}$^*$ & PBEsol DB & - \\
\ce{V6O13}($^*$) & Materials Project & mp-870306 & \ce{Sn2S3}$^*$ & Materials Project & mp-1509 \\
\ce{V7O3}($^*$) & Materials Project & mp-1100921 & \ce{Sn3S} & Materials Project & mp-1187038  \\
\ce{V7O4} & Materials Project & mp-1205504 & \ce{Sn3S4}$^*$ & ICSD & Collection Code 431854 \\
\ce{V7O13}$^*$ & Materials Project & mp-27151 & \ce{Sn3S7}$^*$ & Materials Project & mp-1202007 \\
\ce{V8O}$^*$ & Materials Project & mp-714972 & \ce{CuP2}$^*$ & PBEsol DB & - \\
\ce{V8O15}$^*$ & Materials Project & mp-556566 & \ce{CuP10}$^*$ & Materials Project & mp-606644 \\
\ce{V9O13} & Materials Project & mp-1045898 & \ce{Cu2P} & PBEsol DB & - \\
\ce{V9O17}($^*$) & Materials Project & mp-716723 & \ce{Cu2P3} & PBEsol DB & - \\
\ce{V9O20} & Materials Project & mp-1100929 & \ce{Cu2P7}$^*$ & Materials Project & mp-28034 \\
\ce{V9O22} & Materials Project & mp-849581 & \ce{Cu3P}$^*$ & PBEsol DB & -\\
\ce{V10O9} & Materials Project & mp-705541 & \ce{Cu4P} & PBEsol DB & -\\
\ce{V13O16}$^*$ & Materials Project & mp-30065 & \ce{Cu5P} & PBEsol DB & -\\
\ce{V16O3}$^*$ & Materials Project & mp-30064 & \ce{Cu7P} & PBEsol DB & - \\
\ce{V25O32} & Materials Project & mp-685164 & \ce{Cu8P} & PBEsol DB & - \\
\hline
\hline
\end{tabular}
\caption{All binary compounds included in the construction of the V-O, Ta-N, Sn-S and Cu-P phase predominance diagrams. Binaries that have been experimentally reported are marked with an asterisk. For each composition, the structure that is lowest in energy is chosen here. In case that is not the experimentally reported structure, the asterisk is given in parentheses. Structures extracted from the Materials Project or the ICSD are given with their corresponding IDs. These structures are relaxed with PBEsol and their resulting total energies are used in the fitting of corrections to the elemental reference phase energies.}
\label{tab:binary_systems_additional_structures}
\end{table}

\begin{figure}[h!]
    \centering
    \includegraphics[width=\linewidth]{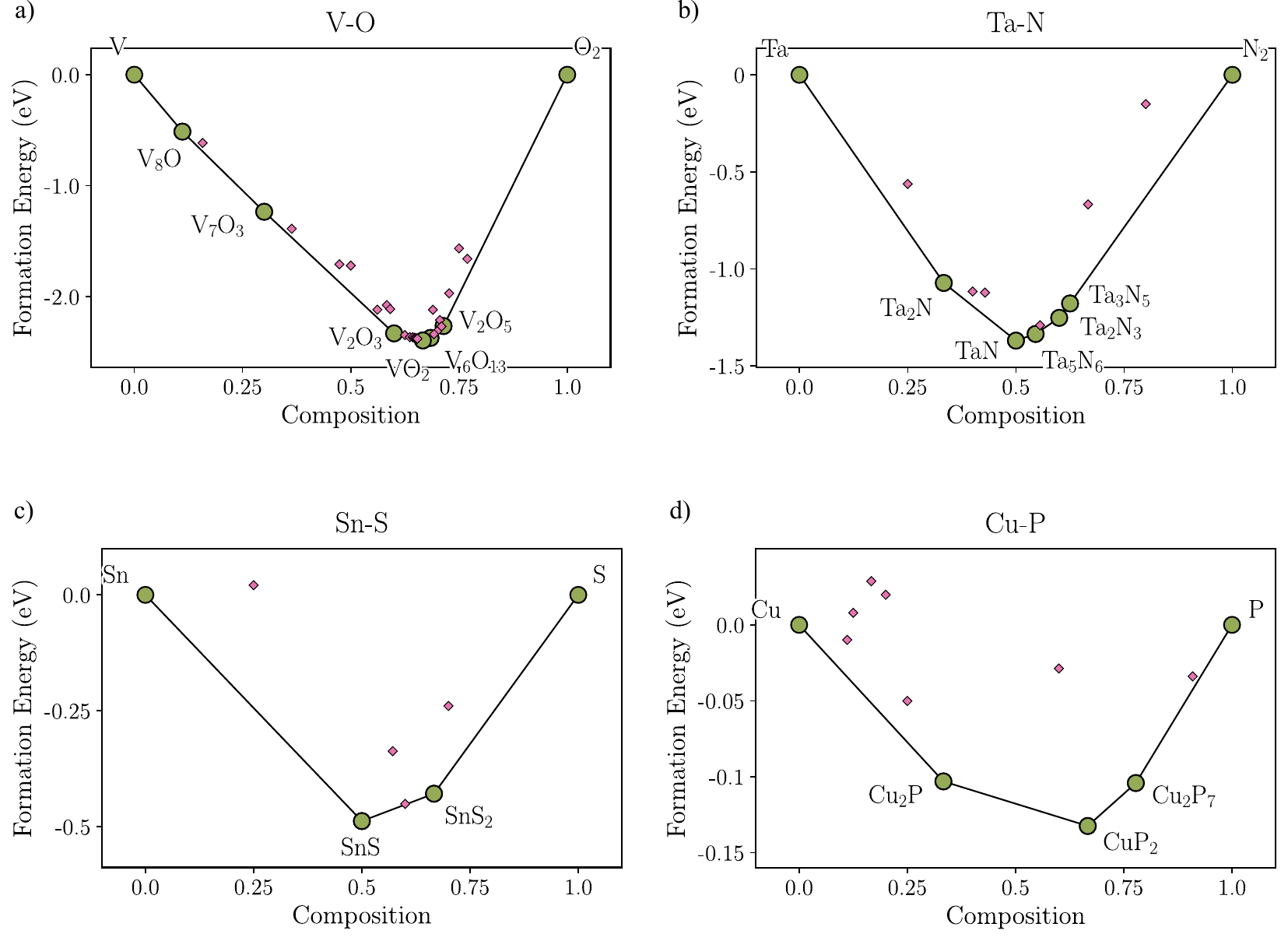}
    \caption{Convex stability hulls calculated at 0~K for the four different binary systems considered in this study.}
    \label{fig:placeholder}
\end{figure}

\newpage

\subsection{Prediction of S and C$_p$ with MatterSim}\label{si:mattersim_prediction}

The entropy S and heat capacity C$_p$ of a compound are derived from its harmonic phonon energy E and the Helmholtz free energy F~\cite{phonopy_1}
\begin{equation}
    S = - \frac{\partial F}{\partial T} = \frac{1}{2T}\sum_{\mathbf{q}\nu}\hbar \omega_{\mathbf{q}\nu} \cdot \coth{\left(\frac{\hbar \omega_{\mathbf{q}\nu}}{2k_B T}\right)} - k_B\cdot \ln{\left[2\sinh{\left(\frac{\hbar \omega_{\mathbf{q}\nu}}{2k_B T}\right)}\right]}
\end{equation}
and
\begin{equation}
    C_V = \left(\frac{\partial E}{\partial T}\right)_V
    = k_B\sum_{\mathbf{q}\nu}\left(\frac{\hbar \omega_{\mathbf{q}\nu}}{k_B T}\right)^2 \cdot \frac{e^{\hbar \omega_{\mathbf{q}\nu}/k_BT}}{(e^{\hbar \omega_{\mathbf{q}\nu}/k_BT} - 1)^2} 
\end{equation}
With $\omega_{\mathbf{q}\nu}$ describing the frequency of the phonon mode with wavevector $\mathbf{q}$ and branch index $\nu$.
The resulting values are normalized per formula unit.
The constant volume heat capacity is derived from the phonon band structure here, however for incompressible solids $C_V = C_p$ holds.
Calculation of both S and C$_p$ is based on a summation over all calculated phonon modes $\mathbf{q}\nu$.  

\subsection{Assessing MatterSim convergence}\label{si:mattersim_convergence}

\begin{figure}[]
    \centering    \includegraphics[width=\linewidth]{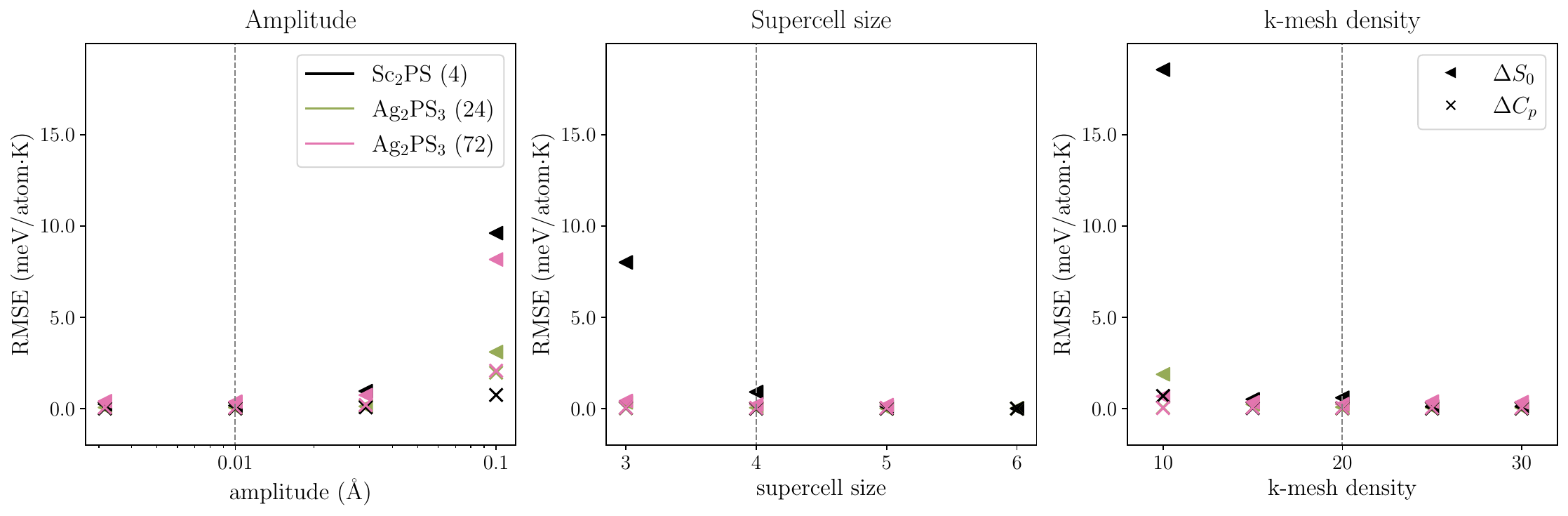}
    \caption{RMSE for both the entropy (triangles) and heat capacity (crosses) that results from varying the amplitude of atomic displacements, the number of unit cells in the supercell, and the density of the k-mesh in the MatterSim calculation. Convergence is assessed for the three indicated phosphosulfide structures containing different numbers of atoms. The values that are selected for the final MatterSim calculation for all materials presented in this study are highlighted with a dashed line.}
    \label{fig:mattersim_convergence}
\end{figure}

To optimize the computational cost of the prediction of entropy $S$ and heat capacity $C_p$ using MatterSim, convergence is assessed by varying the amplitude of atomic displacements, the supercell size, and the k-point grid density. The atomic displacement is varied between 0.001~Å and 0.1~Å in 5 steps. Four different supercell sizes from $2\times2\times 2$ to $5\times 5 \times 5$ are selected. Finally, the number of k-points along one reciprocal lattice direction is varied from 10 to 50. As a metric for convergence, we use the Root Mean Square Error (RMSE) for the temperature dependent entropy and heat capacity over the whole assessed temperature range. The RMSE is determined with respect to the entropy and heat capacity values obtained with the most accurate calculation settings (lowest displacement, largest supercell, densest k-point mesh). The convergence test is performed on three distinct crystal structures with different numbers of atoms in the unit cell. The smallest unit cell for ternary phosphosulfides from the utilized database contains four atoms. For these calculations, \ce{Sc2PS} is selected. The average number of atoms for all structures in this database is 24~atoms, consequently, \ce{Ag2PS3} with a 24~atom unit cell is chosen to represent an average structure from this database. Furthermore, \ce{Ag2PS3} in another configuration with a 72~atom unit cell is chosen to represent structures with a large unit cell from the database. The RMSE resulting from these calculations is shown in Fig.~\ref{fig:mattersim_convergence}. As expected, the RMSE increases with increasing amplitude of atomic displacement and with decreasing supercell size and k-point density. The calculated RMSE is on the same order of magnitude for the entropy and for the heat capacity. When varying the supercell size or the density of the k-mesh, the RMSE is largest for the structure with the smallest unit cell (4 atoms). This is expected, since smaller unit cells correspond to larger Brillouin zones in reciprocal space, requiring denser k-mesh sampling.

To ensure convergence throughout the whole dataset without an unnecessary increase in computational time, 0.01~Å atomic displacement amplitudes, $4\times 4 \times 4$ supercells, and k-meshes of $20\times 20\times 20$ k-points are chosen for all MatterSim calculations in this study. These calculation settings ensure that the resulting RMSE is on the order of $\mu$eV/(atom$\cdot$K) even for the materials with the smallest unit cells. This RMSE is negligible compared to the MAE that results when comparing calculated entropy and heat capacity values with experimental values.

\subsection{Phonon calculations: DFT vs. MatterSim} \label{si:phonon_comparing_DFT_MatterSim}

\begin{figure}[]
    \centering
    \includegraphics[width=.8\linewidth]{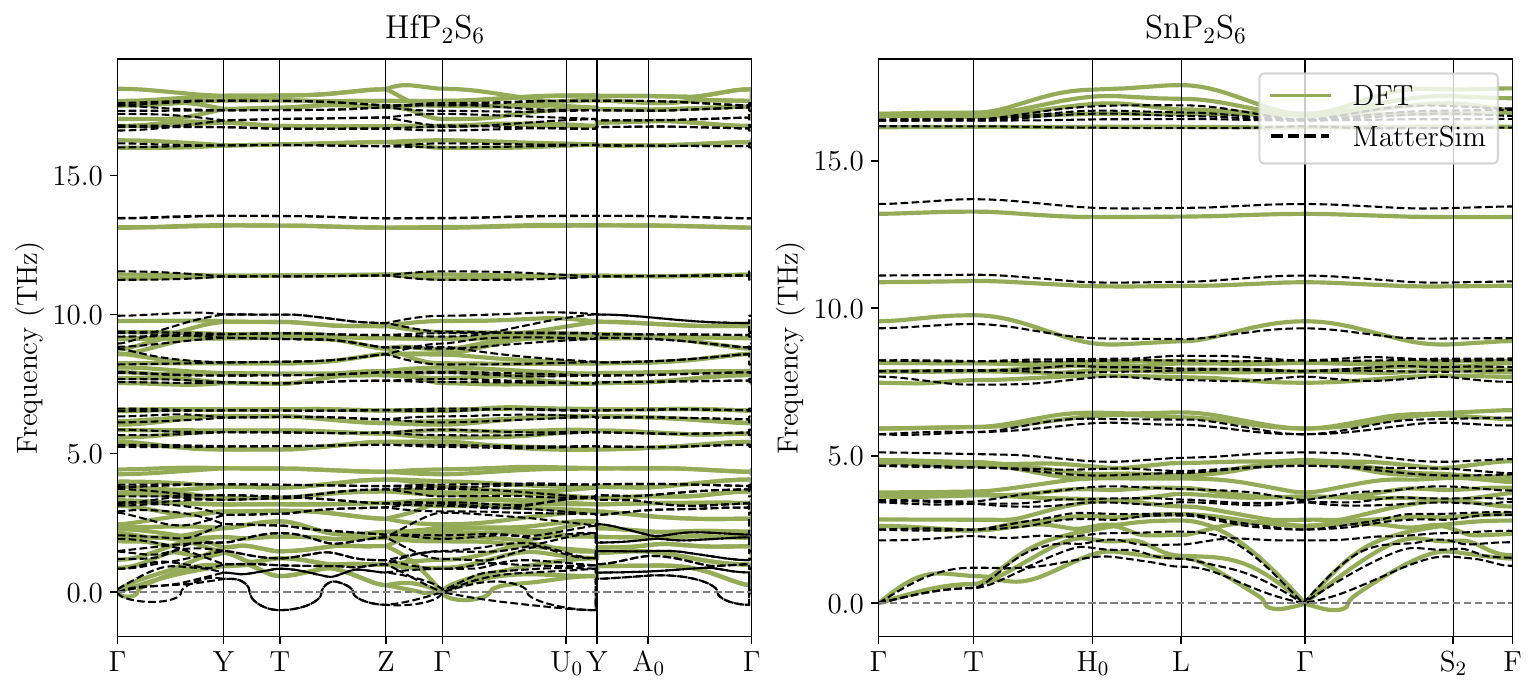}
    \caption{Phonon band structures calculated with DFT~\cite{loewUniversalMachineLearning2025} are compared to the phonon band structures predicted by MatterSim for \ce{HfP2S6} and \ce{SnP2S6}.}
    \label{fig:phonon_bs}
\end{figure}

Phonon calculations are performed by a combination of Phonopy and MatterSim, enabling high-throughput calculations of thermochemical properties. 
MatterSim exhibits a MAE of 0.87~THz on the prediction of the maximum phonon frequency and a MAE of 0.76~THz on the prediction of the average phonon frequency~\cite{MatterSim,loewUniversalMachineLearning2025} with respect to DFT-calculated values with the PBE functional, This is a remarkable performance level for a universal machine-learned interatomic potential model~\cite{loewUniversalMachineLearning2025}. Figure~\ref{fig:phonon_bs} shows an exemplary comparison between phonon band structures of two ternary phosphosulfides (\ce{HfP2S6} and \ce{SnP2S6}) computed from first principles DFT~\cite{loewUniversalMachineLearning2025} and those predicted using MatterSim. Overall, the acoustic and optical branches of the phonon band structures are reproduced by MatterSim with only some exceptions, such as a slight underestimation of the maximum frequency as also reported in MatterSim's original benchmark~\cite{MatterSim}. Although the calculation of imaginary phonon modes may indicate dynamic instability of the material~\cite{imaginary_phonon_frequencies}, both \ce{HfP2S6} and \ce{SnP2S6} have been experimentally reported~\cite{wangSynthesisCrystalStructure1995, simonDarstellungUndAufbau1985}. 

The presence of imaginary phonon modes is assessed for all structures used for the construction of the phase predominance diagrams. All phonon modes exhibiting frequencies below a $-0.3$~THz threshold (standard threshold in MatterSim) are classified as imaginary. Out of the 398 binary structures used in the fitting of the elemental reference phase energies, 79 compounds (20~\% of the total) exhibit imaginary phonon modes in the calculation. Especially in the binary oxide dataset, 47 compounds (31~\%) exhibit imaginary frequencies (binary sulfides: 13 (11~\%), binary pnictides: 12 (12~\%), binary and ternary phosphides and sulfides: 24 (17~\%)). Additionally, 10 (13~\%) of the elemental reference phases are predicted to have imaginary phonon modes. While this indicates potential dynamic instability, several of these structures have been experimentally reported.

Spurious imaginary phonon modes observed in calculated phonon band structures can stem from numerical artifacts, insufficient structural relaxation, or insufficient convergence of the phonon calculation. In such cases, the presence of imaginary modes may be an artifact of the computational setup rather than an indication of true physical dynamic instability of the investigated structure~\cite{imaginary_phonon_frequencies, kamathImpactSpuriousImaginary2026}. It is important to note that these spurious imaginary frequencies can have a significant impact on the prediction of thermodynamical properties like the entropy and heat capacity~\cite{kamathImpactSpuriousImaginary2026}. As a warning to the reader, all compounds that exhibit imaginary frequencies exceeding the threshold of $-0.3$~THz, are marked with a dagger ($^{\dagger}$) in all phase predominance diagrams presented in this work.

\section{Phase predominance diagrams}

\subsection{Comparing experimental and computational phase predominance diagrams}\label{si:error_estimation} 

\subsubsection{Structures and elemental reference energies}

\begin{figure}[h!]
    \centering
    \textbf{Number of compounds = 1} \\ \vspace{0.5cm}
    \includegraphics[width=\linewidth]{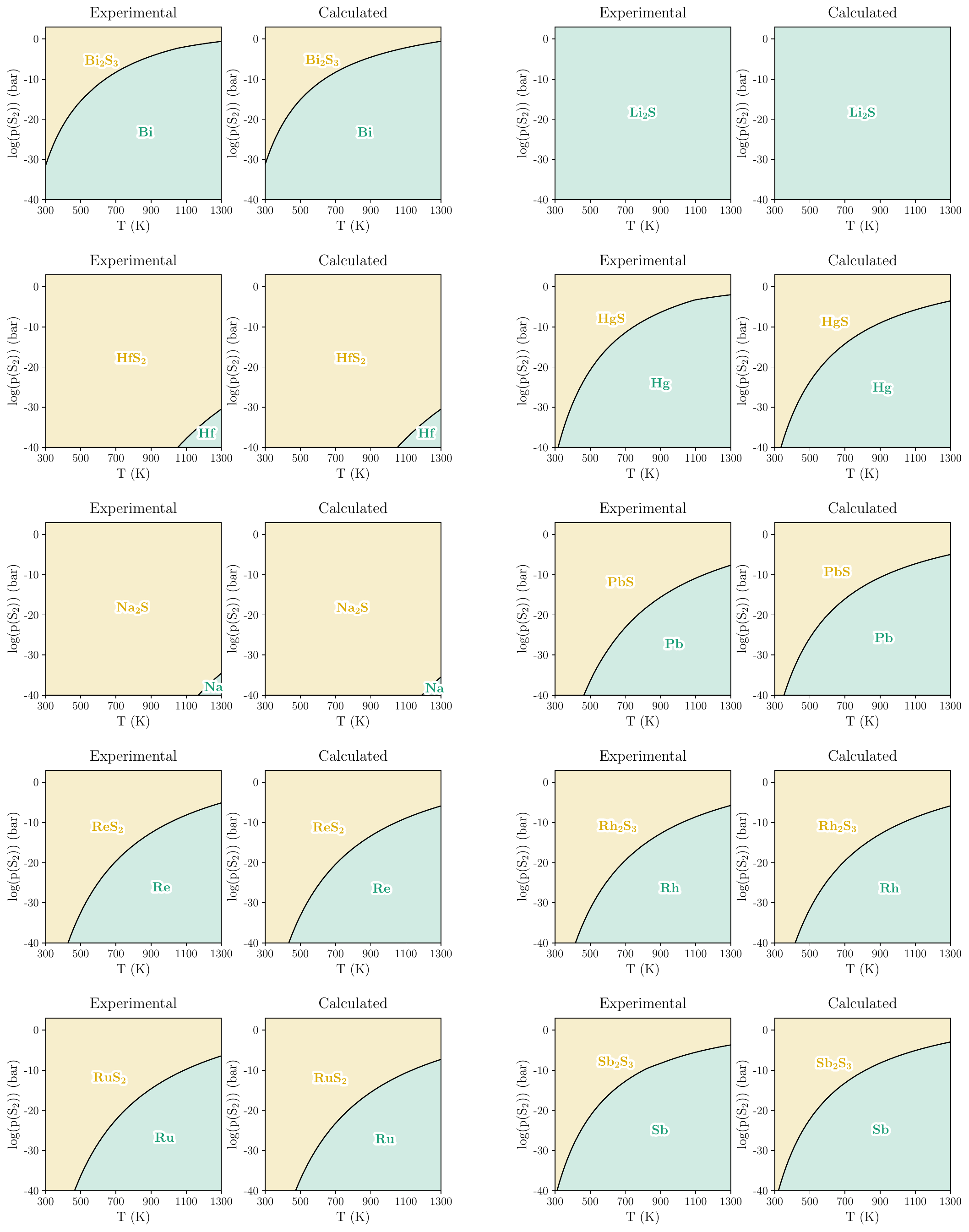}
    \caption{Phase predominance diagrams for the subset of binary systems that contain exactly one binary compound, constructed from experimentally and computationally derived thermodynamic properties. Only the compound that is present in both databases is considered here. The diagrams are presented depending on both temperature and \ce{S2} partial pressure with a fixed \ce{P2} partial pressure of 10$^{-20}$~bar.}
    \label{fig:SI_comparison_exp_calc_1}
\end{figure}

\begin{figure}[h!]
    \centering
    \includegraphics[width=\linewidth]{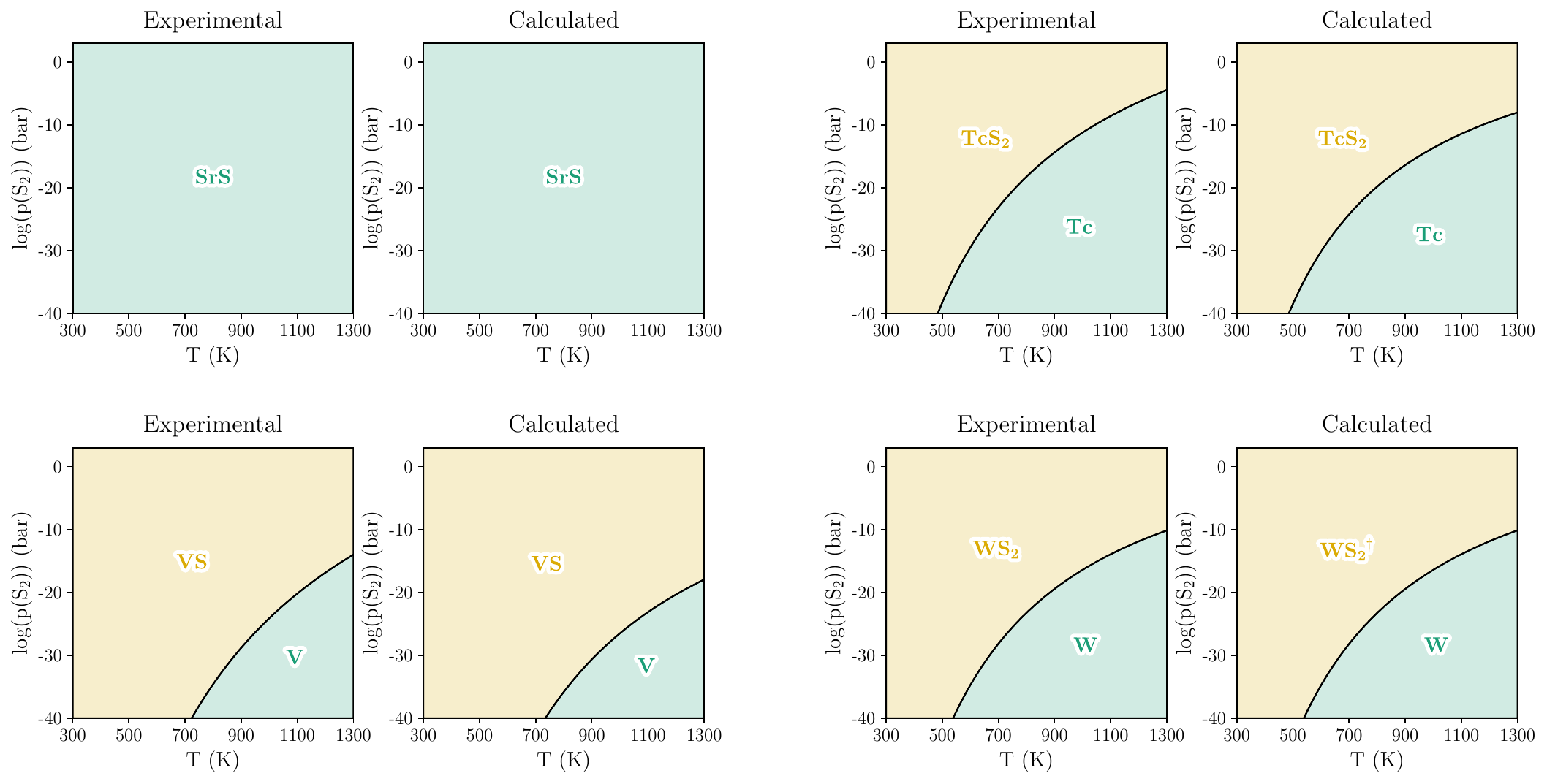}
    \caption{Phase predominance diagrams for the subset of binary systems that contain exactly one binary compound, constructed from experimentally and computationally derived thermodynamic properties. Only the compound that is present in both databases is considered here. The diagrams are presented depending on both temperature and \ce{S2} partial pressure with a fixed \ce{P2} partial pressure of 10$^{-20}$~bar.}
    \label{fig:SI_comparison_exp_calc_2}
\end{figure}

\begin{figure}[h!]
    \centering
    \textbf{Number of compounds = 2} \\ \vspace{0.5cm}
    \includegraphics[width=\linewidth]{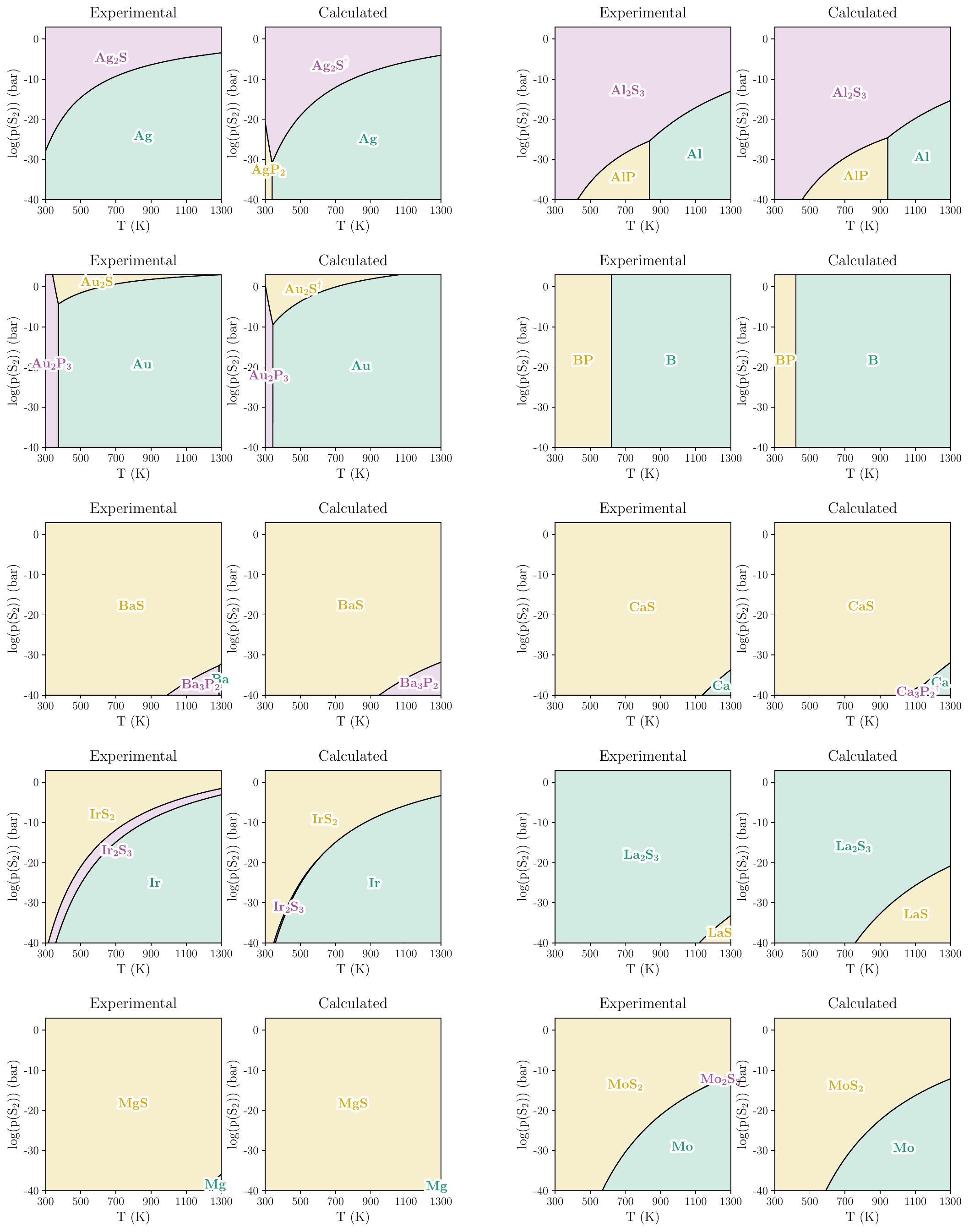}
    \caption{Phase predominance diagrams for the subset of binary systems that contain exactly two binary compounds, constructed from experimentally and computationally derived thermodynamic properties. Only compounds present in both databases are considered here. The diagrams are presented depending on both temperature and \ce{S2} partial pressure with a fixed \ce{P2} partial pressure of 10$^{-20}$~bar.}
    \label{fig:SI_comparison_exp_calc_3}
\end{figure}

\begin{figure}[h!]
    \centering
    \includegraphics[width=\linewidth]{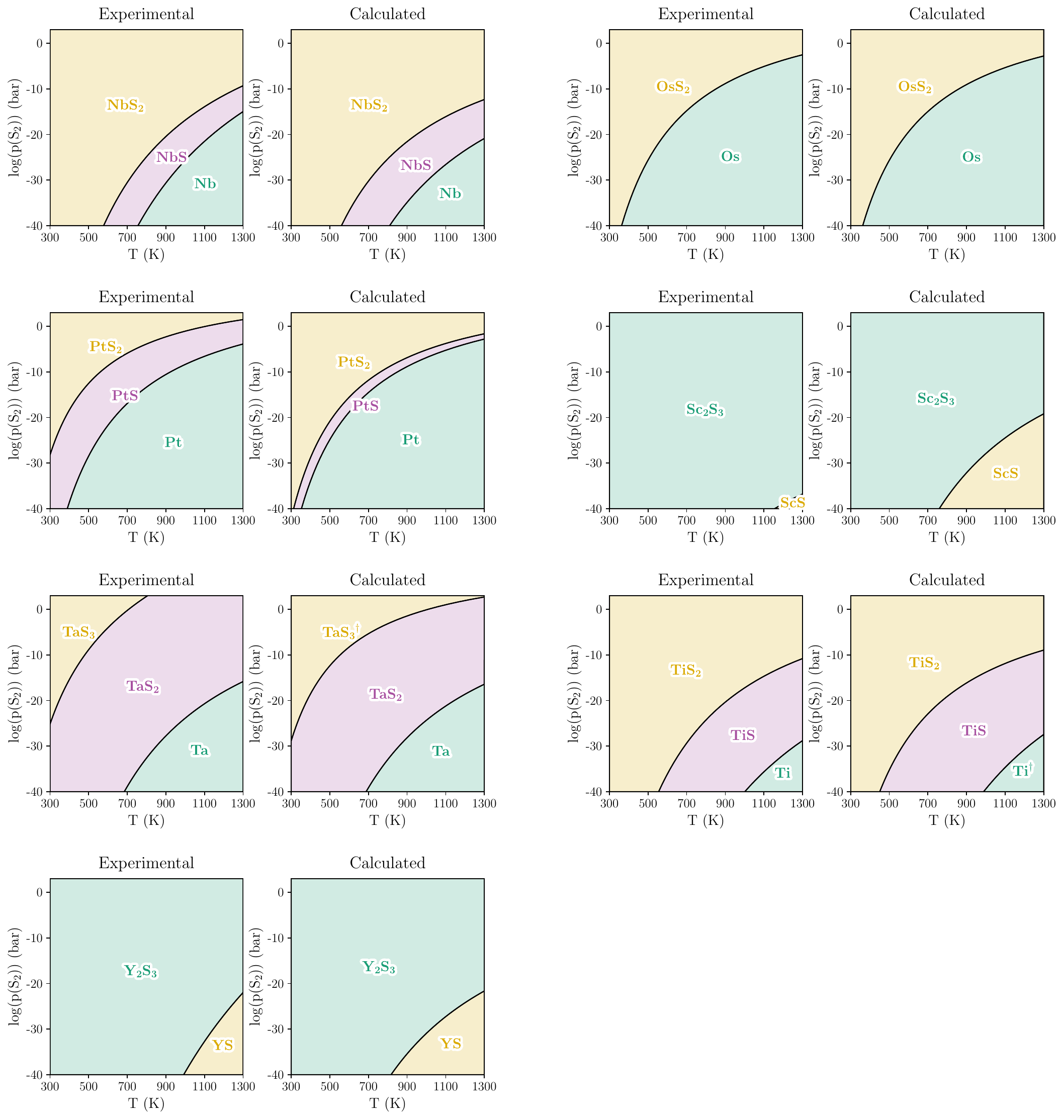}
    \caption{Phase predominance diagrams for the subset of binary systems that contain exactly two binary compounds, constructed from experimentally and computationally derived thermodynamic properties. Only compounds present in both databases are considered here. The diagrams are presented depending on both temperature and \ce{S2} partial pressure with a fixed \ce{P2} partial pressure of 10$^{-20}$~bar.}
    \label{fig:SI_comparison_exp_calc_4}
\end{figure}

\begin{figure}[h!]
    \centering
    \textbf{Number of compounds $\geq$ 3 and $\leq$ 5} \\ \vspace{0.5cm}
    \includegraphics[width=\linewidth]{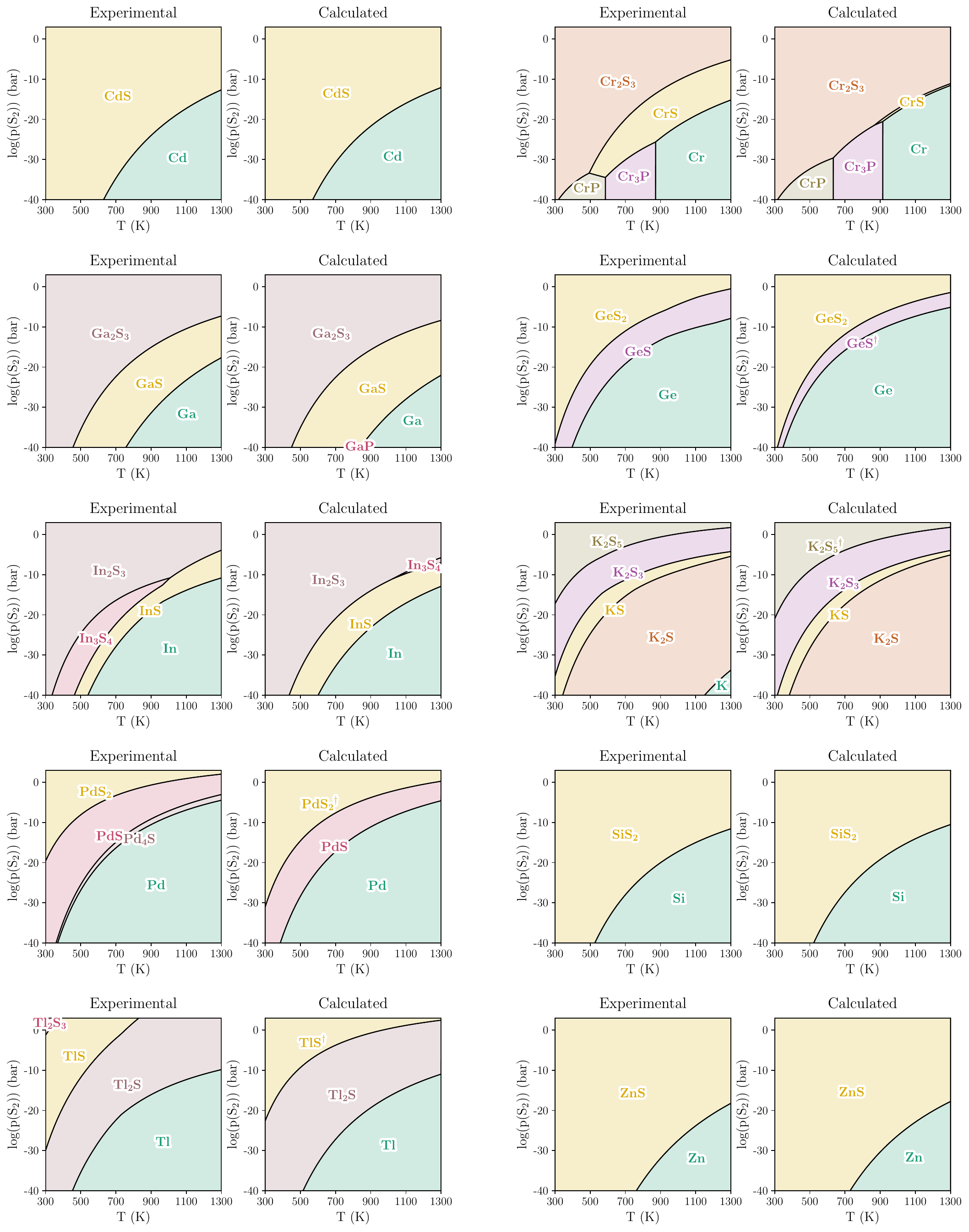}
    \caption{Phase predominance diagrams for the subset of binary systems that contain three to five binary compounds, constructed from experimentally and computationally derived thermodynamic properties. Only compounds present in both databases are considered here. The diagrams are presented depending on both temperature and \ce{S2} partial pressure with a fixed \ce{P2} partial pressure of 10$^{-20}$~bar.}
    \label{fig:SI_comparison_exp_calc_5}
\end{figure}

\begin{figure}[h!]
    \centering
    \textbf{Number of compounds $>$ 5} \\ \vspace{0.5cm}
    \includegraphics[width=\linewidth]{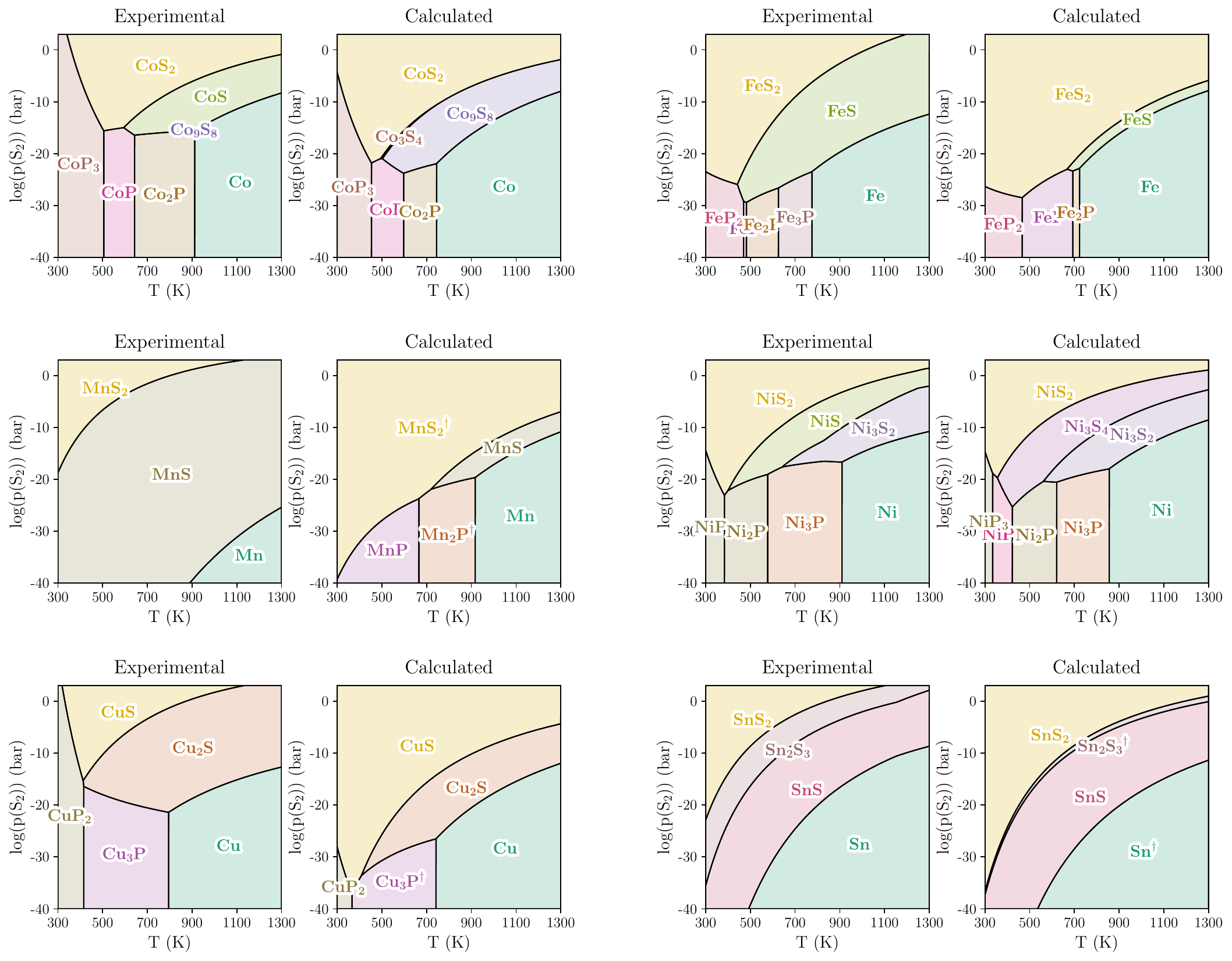}
    \caption{Phase predominance diagrams for the subset of binary systems that contain more than five binary compounds, constructed from experimentally and computationally derived thermodynamic properties. Only compounds present in both databases are considered here. The diagrams are presented depending on both temperature and \ce{S2} partial pressure with a fixed \ce{P2} partial pressure of 10$^{-20}$~bar.}
    \label{fig:SI_comparison_exp_calc_6}
\end{figure}

In the comparison between computational and experimental predominance diagrams, the lowest-energy structure for each composition according to the PBEsol database~\cite{PBEsol_DB} is chosen for the computational diagram. Our reference database of experimental thermochemical properties (HSC Chemistry) usually does not specify the crystal structure of materials, but only their composition. Hence, it cannot be guaranteed that the computed structures always correspond to the structures of the experimentally reported compounds, although this is true in most cases.
~\Cref{fig:SI_comparison_exp_calc_1,fig:SI_comparison_exp_calc_2,fig:SI_comparison_exp_calc_3,fig:SI_comparison_exp_calc_4,fig:SI_comparison_exp_calc_5,fig:SI_comparison_exp_calc_6} show M-P-S phase predominance diagrams constructed based on experimental thermochemical data compared to those derived by the computational workflow presented in this work. The diagrams are grouped by the number of compounds included in the fitting for the elemental reference energy of the corresponding metals. In the first group, consisting of systems where the elemental metal reference energy is fit against the formation energy of a single binary, the nearly perfect agreement between experimental and computational phase boundaries (\Cref{fig:SI_comparison_exp_calc_1,fig:SI_comparison_exp_calc_2}) is essentially due to overfitting. As more compounds are included in the fitting of elemental reference phases, the systems becomes less prone to overfitting. This leads to bigger deviations between the experimental and computational phase boundaries, especially in Fig.~\ref{fig:SI_comparison_exp_calc_6}, but also provides a more representative estimate of the true prediction error in the derived phase boundaries.  

\subsubsection{Estimated error in the temperature of calculated phase boundaries}

The experimentally and computationally derived phase boundaries are compared in the main article in Section~\ref{sec:binaries_comparison_experiment}. To obtain a MAE for the temperature of a calculated phase boundary versus the temperature of the corresponding experimental boundary, we use the following procedure. The temperature difference at constant pressure $\Delta T(p)$ is calculated for each pair of experimental and computational phase boundaries. All phase boundaries that are only present in one of the two diagrams, are excluded from the comparison. The MAE is then calculated via
\begin{equation}
    MAE = \frac{1}{\Delta p} \int_{p_{min}}^{p_{max}} \Delta T (p) dp
\end{equation}
with 
\begin{equation}
    \Delta p = p_{max} - p_{min}
\end{equation}
describing the range of partial pressures where the phase boundary exists.

\FloatBarrier

\subsection{Derivation of  experimental synthesis conditions from literature}\label{si:synthesis_conditions}

Experimental synthesis conditions for ternary phosphosulfides and for binary oxides, phosphides, and sulfides are collected from the literature. Only synthesis procedures based on bulk solid-state reactions in sealed ampoules are considered. No bulk reactions using \ce{N2} as the nitrogen source for the synthesis of Ta-N compounds could be found. The key parameter extracted from each synthesis route is the reaction temperature, from which the partial pressure of gaseous elements can be derived using chemical equilibrium calculations in HSC Chemistry. The initial amounts of elemental phosphorus, sulfur or binary oxides relative to the ampoule volume only have a weak effect on the calculated partial pressures. If the initial quantities of reactants are not reported, stoichiometric amounts are assumed. If the ampoule volume is not reported, a standard volume of \SI{10}{cm^3} is assumed. If neither ampoule volume nor reactant quantities are specified, we tune the stoichiometric masses of the reactants to obtain a total pressure of 10~bar inside the ampoule at the synthesis temperature.

Based on these inputs, the partial pressures of \ce{O2}, \ce{P2} and/or \ce{S2} at the synthesis temperature are calculated using the \textit{Equilibrium Plots} module of HSC Chemistry (Metso)~\cite{HSCChemistry}. This module determines the temperature-dependent partial pressures of all relevant gaseous species. This includes all known elemental oxygen, phosphorus and/or sulfur allotropes, as well as all known P-S binaries if synthesis conditions are derived for M-P-S ternary compounds.

\subsubsection{The Hf-P-S system}

\begin{figure*}[h!]
    \centering
    \includegraphics[width=\linewidth]{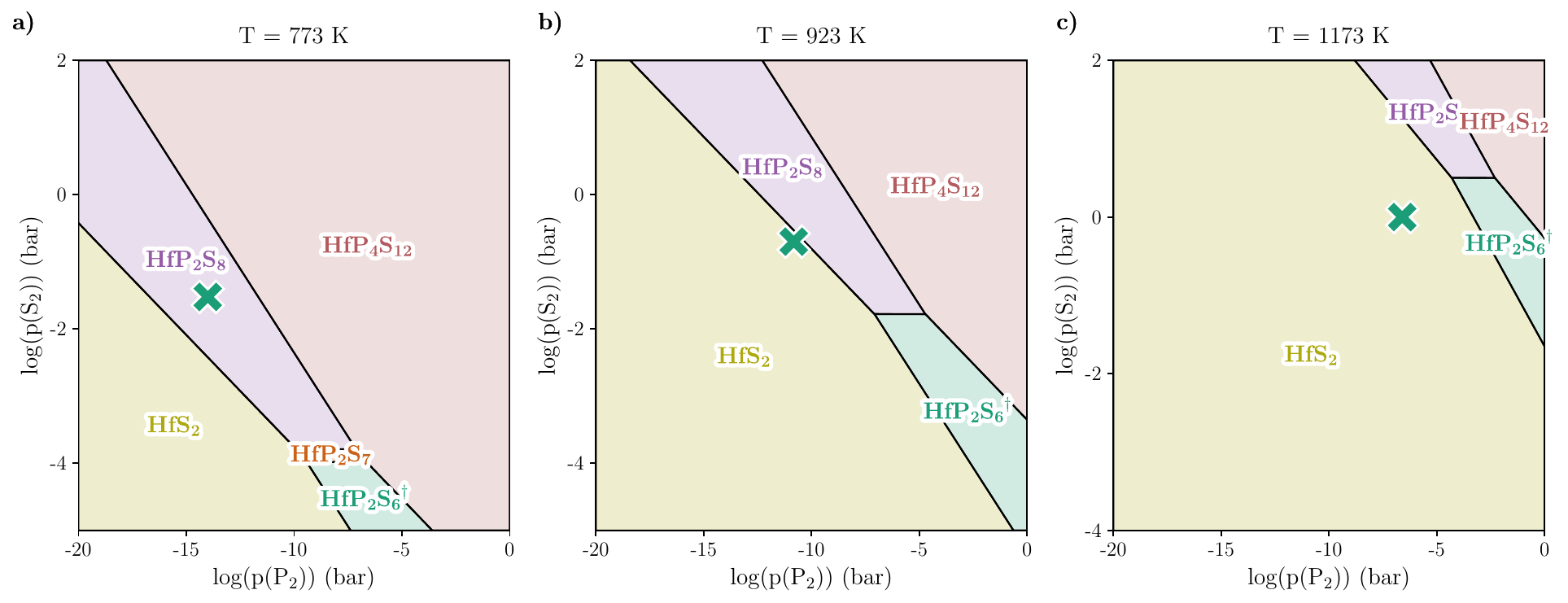}
    \caption{Phase predominance diagrams constructed for the Hf-P-S ternary system. All diagrams are shown in dependence of \ce{P2} and \ce{S2} partial pressures at constant temperature of a) 773~K, b) 923~K and c) 1173~K. Experimental synthesis conditions for \ce{HfP2S6}~\cite{simonDarstellungUndAufbau1985} are marked with a cross in all three diagrams.}
    \label{fig:Hf_P_S_synthesis_conditions}
\end{figure*}

Experimental synthesis conditions for \ce{HfP2S6} are derived from~\cite{simonDarstellungUndAufbau1985} and compared to the phase predominance diagrams for Hf-P-S calculated with the workflow presented in this work in Fig.~\ref{fig:Hf_P_S_synthesis_conditions}. Synthesis of \ce{HfP2S6} has been reported at three different temperatures and the corresponding \ce{P2} and \ce{S2} partial pressures are marked with crosses in Fig.~\ref{fig:Hf_P_S_synthesis_conditions}. At 773~K, the derived synthesis conditions fall within the stability region of the predicted \ce{HfP2S8} compound. At 923~K and 1173~K they lie within the stability region of the experimentally reported binary \ce{HfS2}. With increasing temperature, the derived synthesis conditions move closer to the predicted stability region of \ce{HfP2S6}. Nevertheless, the experimentally reported synthesis conditions for \ce{HfP2S6} are not fully reproduced by the calculated phase predominance diagrams, unlike the case of compounds in the Sn-P-S system. 

\subsection{Full phase predominance diagrams for ternary phosphosulfides}

The method proposed in this work can be scaled to include both more materials, but also complex multinary systems with two gas phase elements. Figures~\ref{fig:ternary_diagrams_1} to \ref{fig:ternary_diagrams_16} show phase predominance diagrams for 48 distinct metal phosphosulfide systems including ternary structures from an in-house, open-access database of phosphosulfides~\cite{Javier_goat}, along with binaries taken from the PBEsol database~\cite{PBEsol_DB}. For each metal, three distinct phase predominance diagram representations are chosen: a phase predominance diagram depending on both \ce{S2} and \ce{P2} partial pressure, calculated at a constant temperature of 800~K, and two temperature dependent phase predominance diagrams that depend on either \ce{S2} or \ce{P2} partial pressure, while the other partial pressure is fixed at \ce{P2} = 10$^{-5}$~bar and \ce{S2} = 10$^{-1}$~bar, respectively.


\begin{figure}
    \centering
    \includegraphics[width=\linewidth]{Figures/ternary_diagrams_0.pdf}
    \caption{Exemplary representations of phase predominance diagrams constructed for the Ag-, Al- and Au-P-S systems. Left: Phase
    diagram for a constant temperature of 800~K in dependence of \ce{S2} and \ce{P2} partial pressures. Middle and right: Phase diagrams dependent
    on temperature and \ce{S2} and \ce{P2} partial pressure, respectively, while the other is kept constant at $10^{-5}$~bar and $10^{-1}$~bar, respectively.}
    \label{fig:ternary_diagrams_1}
\end{figure}

\begin{figure}
    \centering
    \includegraphics[width=\linewidth]{Figures/ternary_diagrams_1.pdf}
    \caption{Exemplary representations of phase predominance diagrams constructed for the B-, Ba- and Bi-P-S systems. Left: Phase
    diagram for a constant temperature of 800~K in dependence of \ce{S2} and \ce{P2} partial pressures. Middle and right: Phase diagrams dependent
    on temperature and \ce{S2} and \ce{P2} partial pressure, respectively, while the other is kept constant at $10^{-5}$~bar and $10^{-1}$~bar, respectively. Features of the B-P-S diagram, such as the very wide range of stability of elemental boron, are likely affected by the absence of boron sulfide binaries from the PBEsol database.}
    \label{fig:ternary_diagrams_2}
\end{figure}

\begin{figure}
    \centering
    \includegraphics[width=\linewidth]{Figures/ternary_diagrams_2.pdf}
    \caption{Exemplary representations of phase predominance diagrams constructed for the Ca-, Cd- and Co-P-S systems. Left: Phase
    diagram for a constant temperature of 800~K in dependence of \ce{S2} and \ce{P2} partial pressures. Middle and right: Phase diagrams dependent
    on temperature and \ce{S2} and \ce{P2} partial pressure, respectively, while the other is kept constant at $10^{-5}$~bar and $10^{-1}$~bar, respectively.}
    \label{fig:ternary_diagrams_3}
\end{figure}

\begin{figure}
    \centering
    \includegraphics[width=\linewidth]{Figures/ternary_diagrams_3.pdf}
    \caption{Exemplary representations of phase predominance diagrams constructed for the Cr-, Cu- and Fe-P-S systems. Left: Phase
    diagram for a constant temperature of 800~K in dependence of \ce{S2} and \ce{P2} partial pressures. Middle and right: Phase diagrams dependent
    on temperature and \ce{S2} and \ce{P2} partial pressure, respectively, while the other is kept constant at $10^{-5}$~bar and $10^{-1}$~bar, respectively.}
    \label{fig:ternary_diagrams_4}
\end{figure}

\begin{figure}
    \centering
    \includegraphics[width=\linewidth]{Figures/ternary_diagrams_4.pdf}
    \caption{Exemplary representations of phase predominance diagrams constructed for the Ga-, Ge- and Hf-P-S systems. Left: Phase
    diagram for a constant temperature of 800~K in dependence of \ce{S2} and \ce{P2} partial pressures. Middle and right: Phase diagrams dependent
    on temperature and \ce{S2} and \ce{P2} partial pressure, respectively, while the other is kept constant at $10^{-5}$~bar and $10^{-1}$~bar, respectively.}
    \label{fig:ternary_diagrams_5}
\end{figure}

\begin{figure}
    \centering
    \includegraphics[width=\linewidth]{Figures/ternary_diagrams_5.pdf}
    \caption{Exemplary representations of phase predominance diagrams constructed for the Hg-, In- and Ir-P-S systems. Left: Phase
    diagram for a constant temperature of 800~K in dependence of \ce{S2} and \ce{P2} partial pressures. Middle and right: Phase diagrams dependent
    on temperature and \ce{S2} and \ce{P2} partial pressure, respectively, while the other is kept constant at $10^{-5}$~bar and $10^{-1}$~bar, respectively.}
    \label{fig:ternary_diagrams_6}
\end{figure}

\begin{figure}
    \centering
    \includegraphics[width=\linewidth]{Figures/ternary_diagrams_6.pdf}
    \caption{Exemplary representations of phase predominance diagrams constructed for the K-, La- and Li-P-S systems. Left: Phase
    diagram for a constant temperature of 800~K in dependence of \ce{S2} and \ce{P2} partial pressures. Middle and right: Phase diagrams dependent
    on temperature and \ce{S2} and \ce{P2} partial pressure, respectively, while the other is kept constant at $10^{-5}$~bar and $10^{-1}$~bar, respectively.}
    \label{fig:ternary_diagrams_7}
\end{figure}

\begin{figure}
    \centering
    \includegraphics[width=\linewidth]{Figures/ternary_diagrams_7.pdf}
    \caption{Exemplary representations of phase predominance diagrams constructed for the Mg-, Mn- and Mo-P-S systems. Left: Phase
    diagram for a constant temperature of 800~K in dependence of \ce{S2} and \ce{P2} partial pressures. Middle and right: Phase diagrams dependent
    on temperature and \ce{S2} and \ce{P2} partial pressure, respectively, while the other is kept constant at $10^{-5}$~bar and $10^{-1}$~bar, respectively.}
    \label{fig:ternary_diagrams_8}
\end{figure}

\begin{figure}
    \centering
    \includegraphics[width=\linewidth]{Figures/ternary_diagrams_8.pdf}
    \caption{Exemplary representations of phase predominance diagrams constructed for the Na-, Nb- and Ni-P-S systems. Left: Phase
    diagram for a constant temperature of 800~K in dependence of \ce{S2} and \ce{P2} partial pressures. Middle and right: Phase diagrams dependent
    on temperature and \ce{S2} and \ce{P2} partial pressure, respectively, while the other is kept constant at $10^{-5}$~bar and $10^{-1}$~bar, respectively.}
    \label{fig:ternary_diagrams_9}
\end{figure}

\begin{figure}
    \centering
    \includegraphics[width=\linewidth]{Figures/ternary_diagrams_9.pdf}
    \caption{Exemplary representations of phase predominance diagrams constructed for the Os-, Pb- and Pd-P-S systems. Left: Phase
    diagram for a constant temperature of 800~K in dependence of \ce{S2} and \ce{P2} partial pressures. Middle and right: Phase diagrams dependent
    on temperature and \ce{S2} and \ce{P2} partial pressure, respectively, while the other is kept constant at $10^{-5}$~bar and $10^{-1}$~bar, respectively.}
    \label{fig:ternary_diagrams_10}
\end{figure}

\begin{figure}
    \centering
    \includegraphics[width=\linewidth]{Figures/ternary_diagrams_10.pdf}
    \caption{Exemplary representations of phase predominance diagrams constructed for the Pt-, Re- and Rh-P-S systems. Left: Phase
    diagram for a constant temperature of 800~K in dependence of \ce{S2} and \ce{P2} partial pressures. Middle and right: Phase diagrams dependent
    on temperature and \ce{S2} and \ce{P2} partial pressure, respectively, while the other is kept constant at $10^{-5}$~bar and $10^{-1}$~bar, respectively.}
    \label{fig:ternary_diagrams_11}
\end{figure}

\begin{figure}
    \centering
    \includegraphics[width=\linewidth]{Figures/ternary_diagrams_11.pdf}
    \caption{Exemplary representations of phase predominance diagrams constructed for the Ru-, Sb- and Sc-P-S systems. Left: Phase
    diagram for a constant temperature of 800~K in dependence of \ce{S2} and \ce{P2} partial pressures. Middle and right: Phase diagrams dependent
    on temperature and \ce{S2} and \ce{P2} partial pressure, respectively, while the other is kept constant at $10^{-5}$~bar and $10^{-1}$~bar, respectively.}
    \label{fig:ternary_diagrams_12}
\end{figure}

\begin{figure}
    \centering
    \includegraphics[width=\linewidth]{Figures/ternary_diagrams_12.pdf}
    \caption{Exemplary representations of phase predominance diagrams constructed for the Si-, Sn- and Sr-P-S systems. Left: Phase
    diagram for a constant temperature of 800~K in dependence of \ce{S2} and \ce{P2} partial pressures. Middle and right: Phase diagrams dependent
    on temperature and \ce{S2} and \ce{P2} partial pressure, respectively, while the other is kept constant at $10^{-5}$~bar and $10^{-1}$~bar, respectively.}
    \label{fig:ternary_diagrams_13}
\end{figure}

\begin{figure}
    \centering
    \includegraphics[width=\linewidth]{Figures/ternary_diagrams_13.pdf}
    \caption{Exemplary representations of phase predominance diagrams constructed for the Ta-, Tc- and Ti-P-S systems. Left: Phase
    diagram for a constant temperature of 800~K in dependence of \ce{S2} and \ce{P2} partial pressures. Middle and right: Phase diagrams dependent
    on temperature and \ce{S2} and \ce{P2} partial pressure, respectively, while the other is kept constant at $10^{-5}$~bar and $10^{-1}$~bar, respectively.}
    \label{fig:ternary_diagrams_14}
\end{figure}

\begin{figure}
    \centering
    \includegraphics[width=\linewidth]{Figures/ternary_diagrams_14.pdf}
    \caption{Exemplary representations of phase predominance diagrams constructed for the Tl-, V- and W-P-S systems. Left: Phase
    diagram for a constant temperature of 800~K in dependence of \ce{S2} and \ce{P2} partial pressures. Middle and right: Phase diagrams dependent
    on temperature and \ce{S2} and \ce{P2} partial pressure, respectively, while the other is kept constant at $10^{-5}$~bar and $10^{-1}$~bar, respectively.}
    \label{fig:ternary_diagrams_15}
\end{figure}

\begin{figure}
    \centering
    \includegraphics[width=\linewidth]{Figures/ternary_diagrams_15.pdf}
    \caption{Exemplary representations of phase predominance diagrams constructed for the Y-, Zn- and Zr-P-S systems. Left: Phase
    diagram for a constant temperature of 800~K in dependence of \ce{S2} and \ce{P2} partial pressures. Middle and right: Phase diagrams dependent
    on temperature and \ce{S2} and \ce{P2} partial pressure, respectively, while the other is kept constant at $10^{-5}$~bar and $10^{-1}$~bar, respectively.}
    \label{fig:ternary_diagrams_16}
\end{figure}

\end{document}